\documentclass[12pt]{iopart}
\usepackage{amssymb}
\usepackage{amstext}
\usepackage{xcolor}
\usepackage{graphicx}

\newcommand{\ket}[1]{\left\vert #1\,\right\rangle}

\newcommand{\braket}[2]{\left\langle #1\,\vert #2\,\right\rangle}

\begin{document}
\title[Macroscopic coherence ]{Macroscopic coherence  as an emergent property in  molecular nanotubes}


\author{Marco Gull\`i}
\address{$^1$Dipartimento di Matematica e
  Fisica and Interdisciplinary Laboratories for Advanced Materials Physics,
  Universit\`a Cattolica del Sacro Cuore, via Musei 41, I-25121 Brescia, Italy}
 \author{Alessia Valzelli}
\address{$^1$Dipartimento di Matematica e
  Fisica and Interdisciplinary Laboratories for Advanced Materials Physics,
  Universit\`a Cattolica del Sacro Cuore, via Musei 41, I-25121 Brescia, Italy}

\author{Francesco Mattiotti}
\address{$^1$Dipartimento di Matematica e
  Fisica and Interdisciplinary Laboratories for Advanced Materials Physics,
  Universit\`a Cattolica del Sacro Cuore, via Musei 41, I-25121 Brescia, Italy}
\address{$^2$Istituto Nazionale di Fisica Nucleare,  Sezione di Pavia, 
  via Bassi 6, I-27100  Pavia, Italy}
\author{Mattia Angeli}
\address{$^1$Dipartimento di Matematica e
  Fisica and Interdisciplinary Laboratories for Advanced Materials Physics,
  Universit\`a Cattolica del Sacro Cuore, via Musei 41, I-25121 Brescia, Italy}
\address{$^3$International School for Advanced Studies (SISSA), Via Bonomea 265, I-34136 Trieste, Italy}
\author{Fausto Borgonovi}
\address{$^1$Dipartimento di Matematica e
  Fisica and Interdisciplinary Laboratories for Advanced Materials Physics,
  Universit\`a Cattolica del Sacro Cuore, via Musei 41, I-25121 Brescia, Italy}
\address{$^2$ Istituto Nazionale di Fisica Nucleare,  Sezione di Pavia, 
  via Bassi 6, I-27100  Pavia, Italy}

\author{Giuseppe Luca Celardo}
\address{$^4$Benem\'erita Universidad Aut\'onoma de Puebla, Apartado Postal J-48, Instituto de F\'isica,  72570, Mexico}

\ead{nicedirac@gmail.com}

\begin{abstract}
Nanotubular molecular self-aggregates are characterized by a high degree of symmetry and they are  fundamental systems for light-harvesting and energy transport. 
While coherent effects are thought to be at the basis of their high efficiency,  the relationship between structure, coherence and functionality is still an open problem. 
We analyze  natural nanotubes  present in  Green Sulfur Bacteria. We show that they have the ability to support macroscopic coherent states, i.e. delocalized excitonic states coherently spread over many molecules, even at room temperature. 
 Specifically, assuming a canonical thermal state, in natural structures we find a
large thermal coherence length, of the order of 1000 molecules.
By comparing natural structure with other mathematical models, we show that
this macroscopic coherence cannot be explained either by the magnitude of the nearest-neighbour coupling between the molecules, which would
induce a thermal coherence length of the order of 10 molecules, or  by the presence of long-range interactions between the molecules. 
Indeed we prove that the existence of macroscopic coherent states is  an emergent property of such structures due to  the interplay between geometry and cooperativity (superradiance and super-transfer).
In order to prove this, we give evidence that the lowest part of the spectrum of natural systems is determined by a cooperatively enhanced coupling (super-transfer) between  the  eigenstates of modular sub-units of the whole structure. 
Due to this enhanced coupling strength,  the density of states is lowered  close to the ground state, thus boosting the thermal coherence length.
As a striking consequence of the lower density of states,  an energy gap between the excitonic ground state and the first excited state emerges. Such energy gap   increases  with the length of the nanotube (instead of decreasing as one would expect), up to a critical system size which is close to the  length of the natural complexes considered.

\end{abstract}



\section{Introduction}
Coherent effects, as fragile as they may seem, might be able to survive in complex 
systems even in presence of strong noise induced by the coupling to an external environment. 
They are often related to functions in complex chemical and biophysical systems~\cite{photo,photoT,coherence}. 
Understanding under which  conditions robust coherent effects can be sustained even at room temperature  is a central issue for designing efficient quantum devices.

Molecular nanotubes are among the most interesting and most investigated structures. They are present in several natural photosynthetic complexes, for instance in the Green Sulphur Bacteria~\cite{Valleau,Huh,arvi1,arvi2,arvi3,K31,K32} or in Phycobilisome Antennas~\cite{nir1,nir2,nir3,nir4}. They are also present in other biomolecular systems, for instance in Microtubules, which are fundamental biological structures, showing interesting similarities with photosynthetic Antenna complexes~\cite{craddock,Phil}. 
Also artificial  molecular nanotubes  are  at the centre of research interest~\cite{caoNT,eisele,NT2,K1}. 
Nanotubular molecular aggregates are extremely efficient for light-harvesting and energy transport and they present a very ordered structure with a high degree of  symmetry~\cite{arvi1,arvi2,arvi3,K2,K41,K42,K43}. 
The high degree of symmetry concerns both
  the molecule positions and the orientation of their transition dipoles.
Despite all that,
a clear  understanding of  how structural features in  molecular aggregates can  sustain coherent effects and  explain their high efficiency is still missing. 

Some of the primary coherent effects which are thought to be responsible
for the high efficiency of molecular nanotubes are
induced by the delocalization of the excitation over many 
molecules. Since the sunlight is very dilute, usually only one excitation is present in such complexes, so that single-excitation delocalized states are   usually investigated. Delocalized excitonic states 
can lead to cooperative effects, such as superradiance~\cite{K31,K32,K2,Jaggr,Moll,vangrondelle}
and super-transfer~\cite{schulten,srlloyd}, and they can be useful 
in both natural or artificial 
light-harvesting complexes~\cite{eisele,vangrondelle,superabsorb,kaplan1,kaplan2,kaplan3,kaplan4,kaplan5,kaplan6,kaplan7,kaplan8,sr2,srfmo,srrc}. 
Specifically, coherently delocalized excitonic states can have a 
large dipole strength which strongly couples them to
the electromagnetic field. Thus, these states are able to
super-absorb light  at  a rate much larger than the single-molecule absorbing rate, since
 the absorption rate of delocalized excitonic states can
increase with the number of molecules over which the excitation is
delocalized~\cite{K31,K32}. States with a large dipole strength can also couple between themselves
efficiently, inducing a super-transfer coupling between distant molecular aggregates~\cite{srlloyd} or  different parts of the same aggregate as we show here. 
Delocalized single excitonic states over a large number of molecules are called macroscopic coherent states and they  are studied both for  applications and basic science~\cite{mc1,mc2,Cl,eisele2,eisele3,JYZ,JRC}.

Molecular nanotubes are composed by a  network of self-assembled photo-active molecules. 
Each molecule can be treated as  a two level system, characterized by both an excitation energy and a transition dipole moment which determines its coupling with the electromagnetic field and with the other molecules. 
The interaction between the molecules is often assumed to be  dipole-dipole~\cite{K2,K41,K42,K43} which decays with the distance as $1/r^3$ or,
in some approximate scheme, as 
nearest-neighbour~\cite{K1} only. While the results thus obtained are certainly very interesting, care is needed to use such simplifications in large molecular structures. Indeed, dipole-dipole interaction is valid when the distance between the molecules is sufficiently large and the overall   system
size $L$ is considerably smaller than the wavelength $\lambda_0$ connected with the excitation energy of the molecules (small volume limit).
Since nanotubular aggregates can be large, here we consider a more accurate Hamiltonian interaction~\cite{mukamelspano} which takes into account the interaction between oscillating charges in each molecule. Such description reduces to the usual dipole-dipole interaction in the small volume limit.

Using such radiative Hamiltonian, we have analyzed the existence of macroscopic coherent states at room temperature in different, natural and artificial, molecular nonotubes. 
Since the molecules in such structures are tightly packed, their interaction energy can be strong, of the order of several times $ k_B T \approx 200\mbox{ cm}^{-1}$ with $T=300K$. Such strong interaction is thought to be able to support excitonic delocalization even at room temperature. Nevertheless here we show that the symmetric arrangement of the molecules is able to induce excitonic delocalization at room temperature well beyond what one could expect from the magnitude of the nearest-neighbour coupling between the molecules. Moreover, by comparing natural structures with few mathematical models of self-aggregated molecular nanotubes we show that the degree of macroscopic coherence cannot be explained even by the long-range nature of the coupling between the molecules. We connect such enhanced delocalization to the super-transfer coupling present inside such structures, which induces the emergence of a gapped superradiant state in the low energy region of the spectrum. 
Thus our main result is that
macroscopic coherence in natural molecular nanotubes is  an emergent property produced by specific cooperative effects which cannot be reduced either to the range of the interaction or to the magnitude of the coupling between the molecules.

Specifically, in this paper  we investigate the \textit{Chlorobium Tepidum} Antenna complexes of Green Sulfur bacteria. Green Sulfur bacteria are photosynthetic organisms which live in deep water where the sunlight flux is very low~\cite{Huh} and  they are among the most efficient photosynthetic systems~\cite{arvi1,arvi2,arvi3}.
Similarly to other antenna complexes present in nature~\cite{nir1,nir2,nir3,nir4}, they present a high degree of symmetry being
arranged in nontrivial cylindrical structures with an ordered orientation of the molecule dipoles. 
We analyze both  
the wild type  (WT) and the triple mutant type (MT),  which have been recently investigated in~\cite{Ganapathy,Koh}. 
  
Understanding the connection between functionality and structure in such complexes  will enhance our comprehension  of natural photosynthesis and it could  also inspire efficient bio-mimetic devices for energy transport and light-harvesting.


In Section~\ref{sec:mod} and~\ref{sec:ham} we present the cylindrical models studied. In Section~\ref{sub-41} the existence of a delocalized superradiant state close to the ground state for the natural models is shown. In Sections~\ref{sub-42} and~\ref{sub-43} the thermal coherence length is introduced and analyzed. Natural complexes are shown to be able to support the largest thermal coherence length w.r.t. the other models considered. The evidence produced in these Sections allows to conclude that the large thermal coherence length of natural aggregates cannot be explained by the magnitude of the coupling or by the range of the interaction between the molecules. In Section~\ref{sec:rel} we explain that the origin of such macroscopic coherent states found in natural complexes lies in their specific geometry which induces a supertransfer coupling inside the complexes. Such supertansfer coupling strongly affect the lowest part of the spectrum thus enhancing the thermal coherence length. In Section~\ref{sec:conc},
we analyze structures which are more complex than single cylindrical surfaces. Specifically,  we consider tubular structures made of four concentric cylindrical surfaces, as they appear in natural antenna complexes of Green Sulfur bacteria~\cite{Arellano,Ganapathy,Koh,Chew}.  We show that these structures display an enhanced delocalization of the excitation with respect to single cylindrical surfaces. 
Finally in Section~\ref{conclu} we give our conclusions and perspectives.


\section{The models}
\label{sec:mod}
The natural Antenna complexes present in Green Sulphur bacteria have lengths of $1000$ - $2000$ \AA,  widths of $100$ - $600$ \AA \, and they can contain a number of molecules between $50,000$ and $250,000$, typically arranged into concentric cylindrical surfaces \cite{Huh,Linnanto}. 
It is important to remark that, depending on the environment and on the growing conditions \cite{Hohmann}, some samples could show an alternation between tubular aggregates and non-tubular curved lamellae \cite{Ikonen,Oostergetel}. Nevertheless, in spite of the heterogeneity of the structures experimentally observed, we will consider here cylindrical surfaces only with a radius of $6$ nm and length up to $L=250$~nm  composed of $1500$ molecules.

Specifically,  we   analyse five different
cylindrical models with  fixed radius ($R=60 \mbox{ \AA}$) and  total number of chromophores $N$.   These models differ for  
   the geometrical arrangement of the chromophores along the cylindrical surface. In details they are: 
\begin{itemize}
\item \textit{Chlorobium Tepidum} bchQRU triple mutant (MT),
\item \textit{Chlorobium Tepidum} wild type (WT),
\item parallel dipoles cylinder (PD),
\item tangent dipoles cylinder (TD),
\item random dipoles cylinder (RD). 
\end{itemize}
While the first two are representative of natural systems, the others are mathematical models with a suitable symmetric arrangements of chromophores
(TD and PD) while the last one (RD) is characterized by a random orientation of the dipole moments. The molecule positions and dipole orientations for the natural models have been taken from literature~\cite{Ganapathy,Koh,Chew} and they correspond to the values capable to reproduce experimental results.
\begin{figure}[t] 
\centering
\includegraphics[scale=0.16]{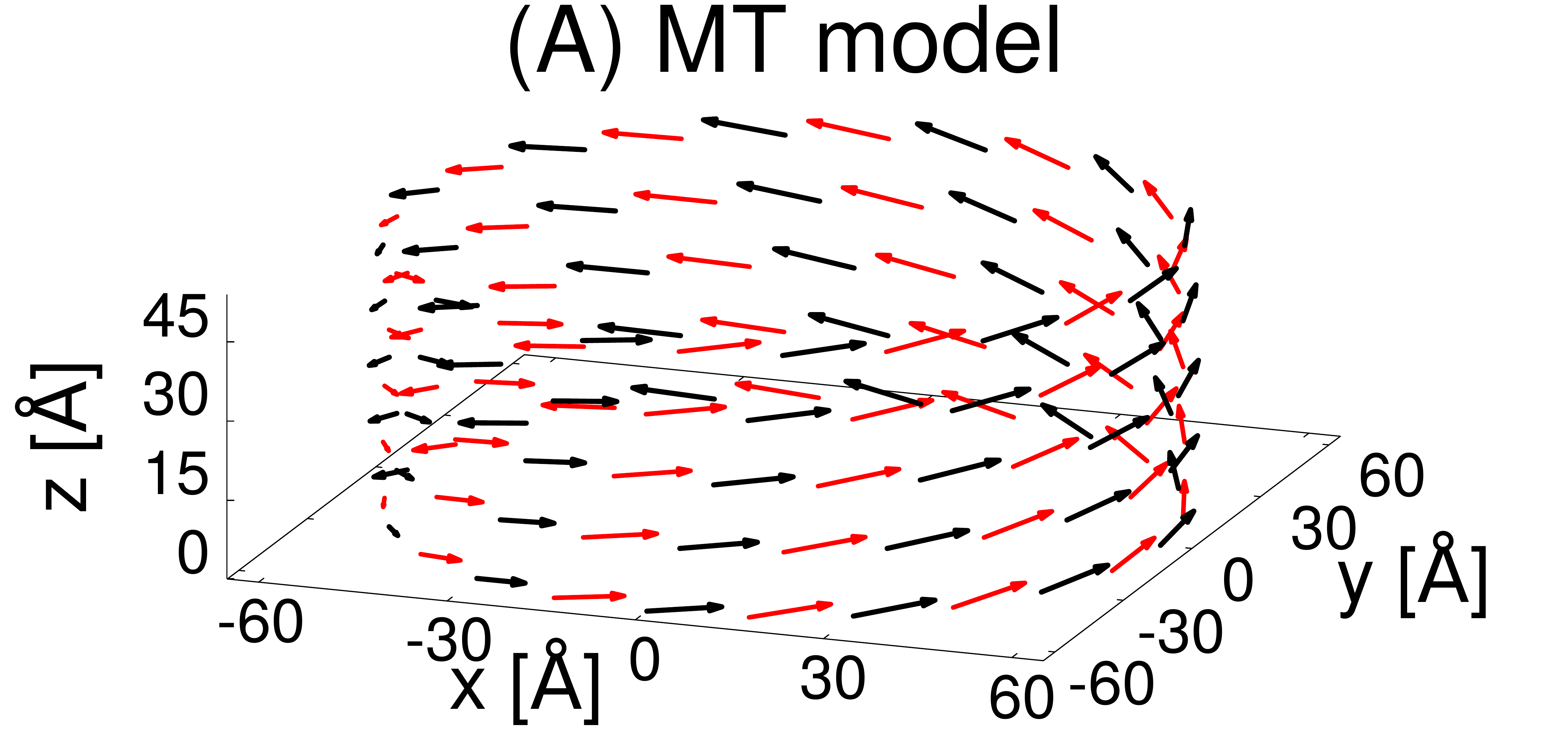}
\includegraphics[scale=0.16]{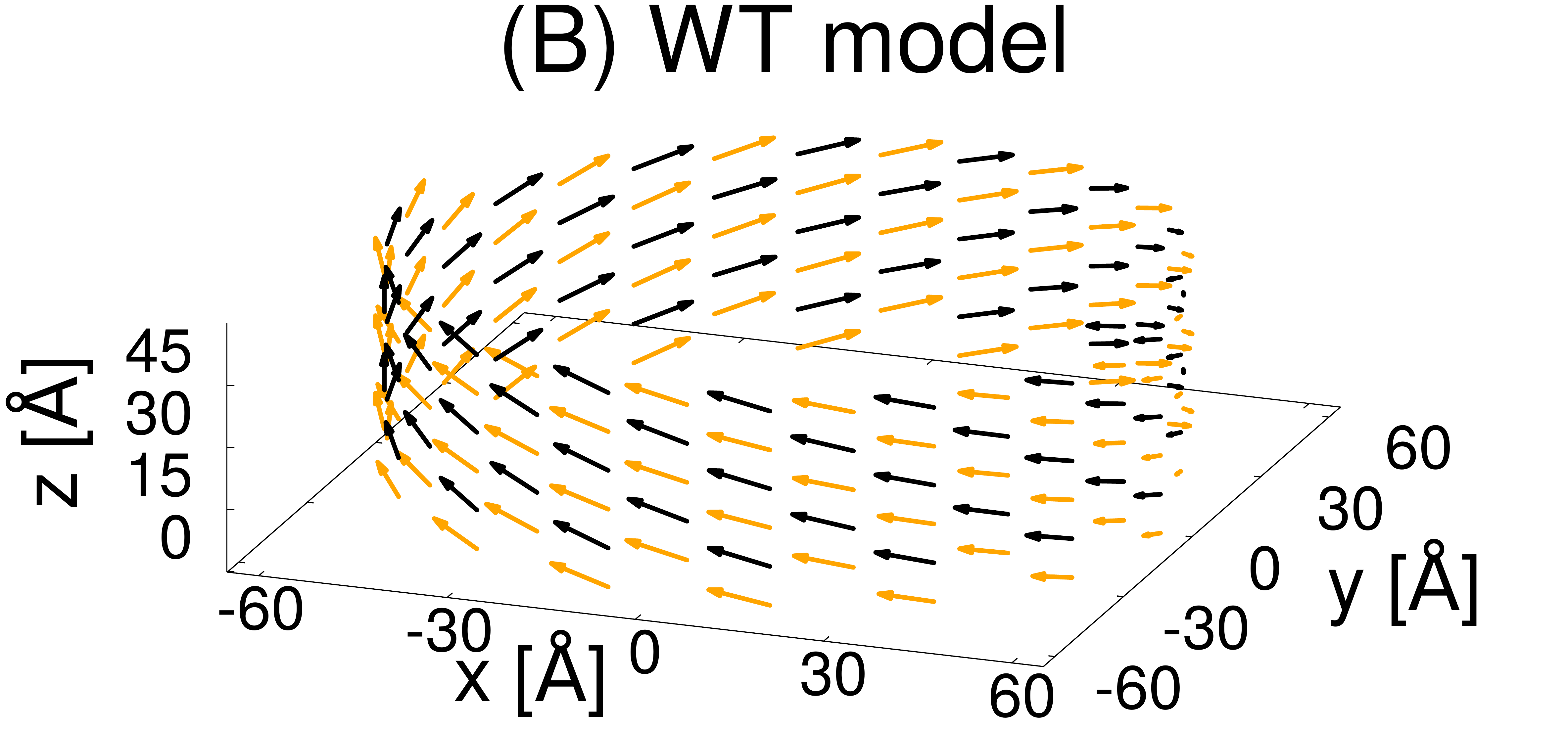}
\includegraphics[scale=0.16]{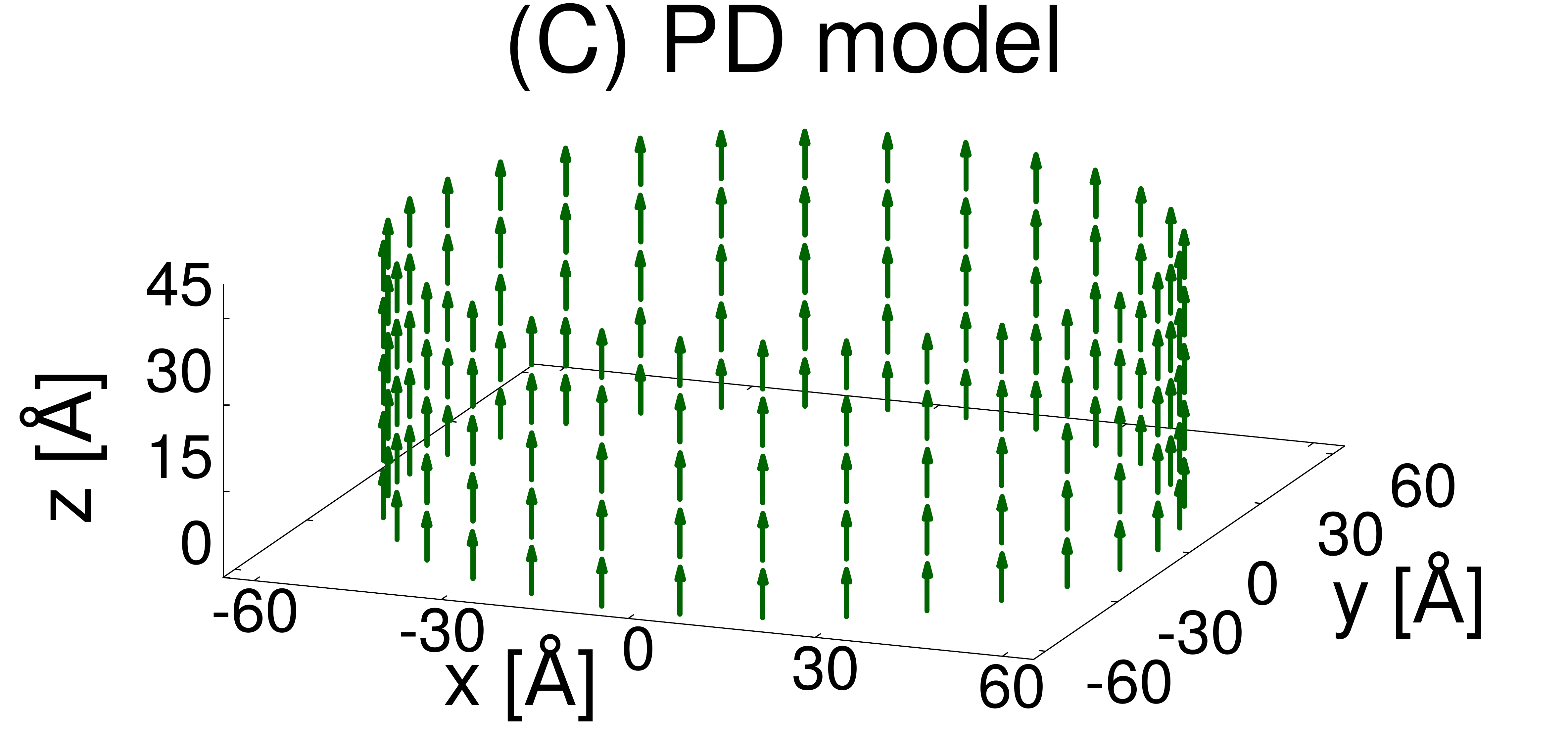}
\includegraphics[scale=0.16]{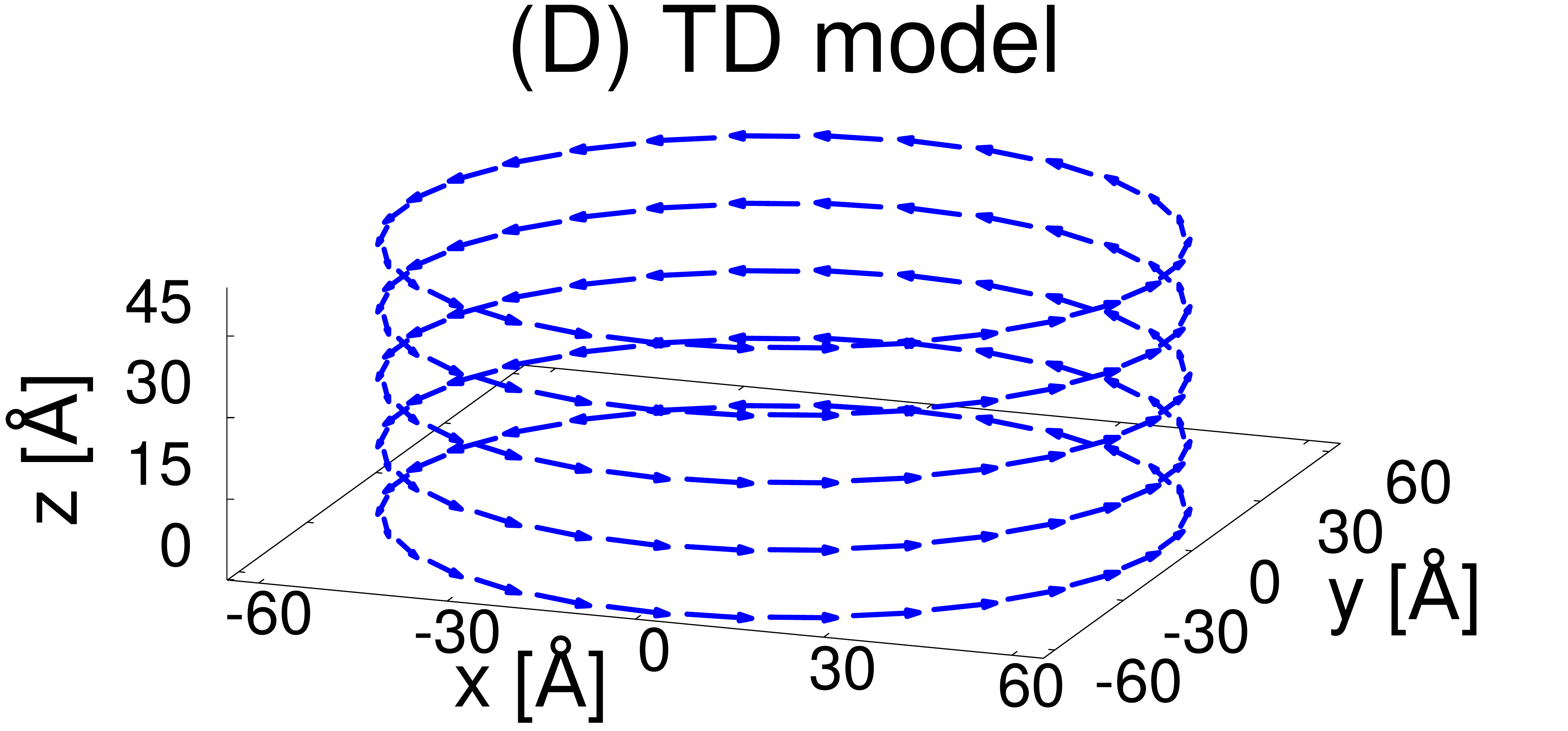}
\includegraphics[scale=0.16]{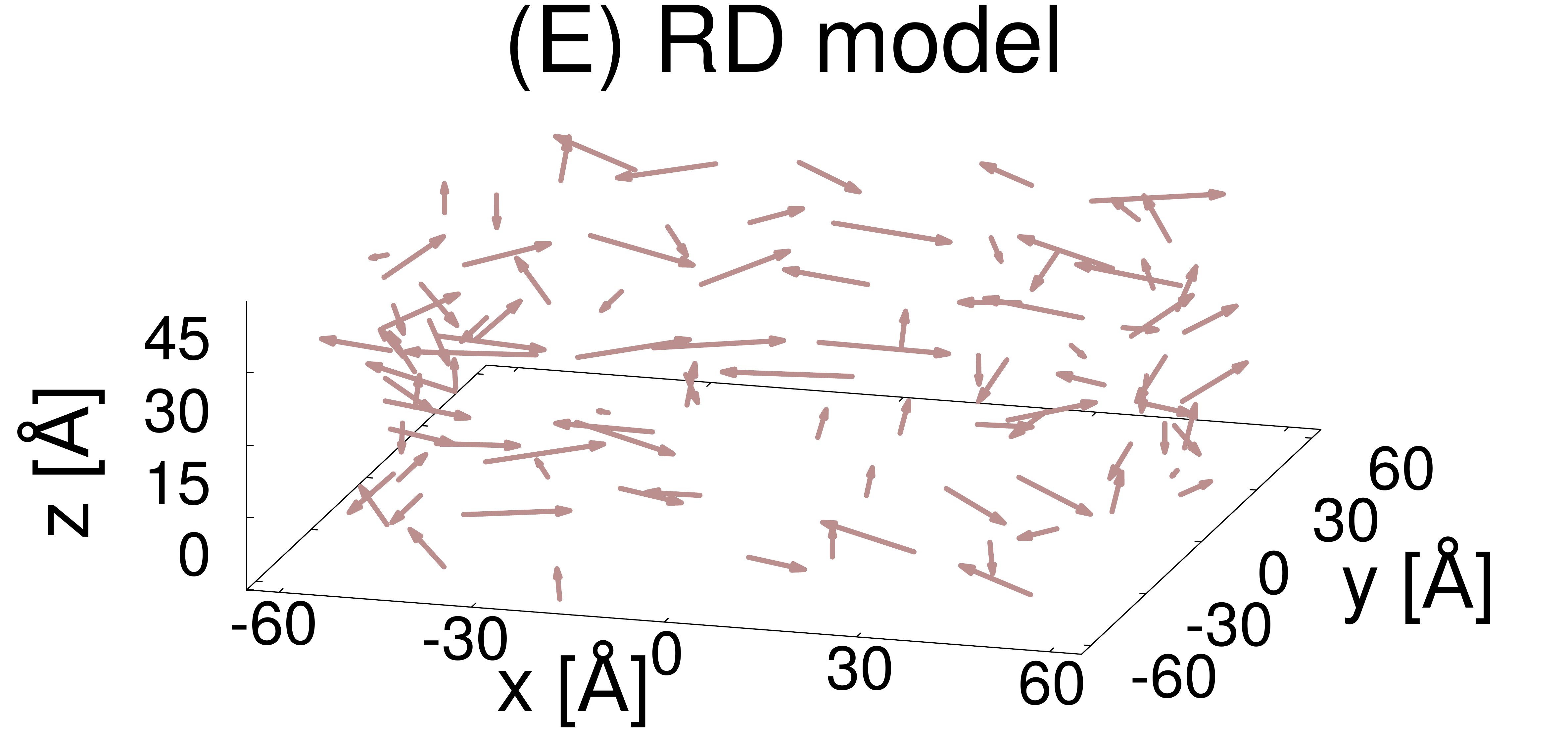}
\caption{Sections of the different models.  
In all panels we show cylinders 
with the same radius $R=60 \mbox{ \AA}$. For the sake of clarity we show only $30$ dipoles per ring instead of $60$ as we considered in this paper.
Moreover the distances along the $z-$axis are enhanced by a factor of 5 with respect to the distances on the $x-y$ axes. 
The same factor and also a reduction of the number of dipoles  have been used for WT model. In all models but the WT, where the dipoles are arranged in a helical structure, the dipoles are arranged into $N_1=5$ rings.  
}
 \label{fig:cyl-all}
\end{figure}
 
A schematic view of the arrangement of the dipoles on the cylindrical surfaces for all models is shown in Figure~\ref{fig:cyl-all}, while all other technical details
can be found in \ref{app-a}. Notice that all the models but the WT share the same basic structure: the cylinder is made by a collection of $N_1$ rings composed of $N_2=60$ molecules equally spaced on each ring.  The difference between them  lies in the dipole orientation only: 
\begin{itemize}
    \item PD model: all dipoles are oriented parallel to the $z$ axis
    \item TD model: all dipole are perpendicular to the $z$ direction and tangential to the cylindrical surface
    \item MT model: here the dipoles have a fixed $z$ component, but also a component perpendicular to the $z$ direction, see~\ref{app-a} for details. Note that the component perpendicular to the $z$ direction points inward and outward alternatively with respect to the plane tangent to the cylindrical surface with a small angle $\alpha$ (see black and red arrows in Figure~\ref{fig:cyl-all}(A)).
    \item RD model: the position of the dipoles is the same of the other three models but the orientation of the dipoles is fully random on the unit sphere.
\end{itemize}
On the other hand the WT model, see Figure~\ref{fig:cyl-all}(B), is not composed of separated rings but instead is arranged in a  complicated helical structure, see~\ref{app-a} for details.

\section{The Hamiltonian and the dipole approximation}
\label{sec:ham}
Each molecule is represented as a two-level system with an excitation energy $e_0$ and a transition dipole moment $\vec{\mu}$. The parameters of the aggregates considered here have been taken from literature~\cite{Holz,Steens} to be the ones characterizing the Antenna Complexes in Green Sulfur bacteria. 
Specifically we set for the excitation energy of all the molecules $e_0 = 15390 \mbox{ cm}^{-1}$~\cite{Steens}, corresponding to $\lambda_0 \approx 650$ nm, so that
\begin{itemize}
    \item[] $k_0=2 \pi e_0 \times 10^{-8}= 9.670 \times 10^{-4} \mbox{  \AA}^{-1}$.
    \item[] $\mu=\sqrt{30} \mbox{ D}$~\cite{Holz} so that $|\mu|^2= 151024 \mbox{  \AA}^3 \mbox{  cm}^{-1}$ (for the conversion, see~\cite{dipsquare}). 
    \item[] $\gamma= 4 |\mu|^2 k_0^3/3= 1.821 \times 10^{-4} \mbox{ cm}^{-1}$, corresponding to the radiative lifetime $\tau_\gamma=29.15\text{ ns}$ (for the conversion, see~\cite{enertime}).
\end{itemize}

Choosing the basis states in  the single excitation manifold, where the state
 $|i\rangle$ refers to a state in which the $i^{th}$ molecule is excited while all the others are in the ground state, the nanotubes can be described 
 through a Non-Hermitian Hamiltonian which takes into account the interaction between the molecules mediated by the electromagnetic field
(EMF).
 The effective Non-Hermitian Hamiltonian (also called radiative Hamiltonian),  is commonly used to model the interaction with the  EMF in different  systems, such as natural light-harvesting
 complexes~\cite{mukamelspano,mukameldeph} and cold atomic clouds~\cite{kaiser} and it reads: 
\begin{equation} \label{eq:ham}
H=\sum_{i=1}^N e_0|i\rangle \langle i|+\sum_{i\neq j}\Delta_{ij}|i\rangle \langle j|-\frac{i}{2}\sum_{i,j=1}^{N}Q_{ij}|i\rangle \langle j|.
\end{equation}
The   terms $\Delta_{ij}$ and $Q_{ij}$ derive from  the interaction with the  EMF.
The real and imaginary diagonal parts of the intermolecular coupling are given   respectively, by
\begin{equation}
  \Delta_{nn} = 0 \, , \\
  Q_{nn} = \frac{4}{3} \mu^2 k_0^3 = \gamma \, , \label{eq:gamma}
\end{equation}
with $\mu=|\vec{\mu}|$ being the transition dipole, while the off-diagonal ($n \ne m$)  by
$$
  \Delta_{nm} = \frac{3\gamma}{4} \left[ \left( -\frac{\cos (k_0 r_{nm})}{(k_0 r_{nm})} +
    \frac{\sin (k_0 r_{nm})}{(k_0 r_{nm})^2} + \frac{\cos (k_0 r_{nm})}{(k_0 r_{nm})^3} \right)
    \hat{\mu}_n \cdot \hat{\mu}_m +\right. \nonumber \\
$$    
\begin{equation}
    -\left. \left( -\frac{\cos (k_0 r_{nm})}{(k_0 r_{nm})} + 3\frac{\sin (k_0 r_{nm})}{(k_0 r_{nm})^2} +
    3\frac{\cos (k_0 r_{nm})}{(k_0 r_{nm})^3}\right) \left( \hat{\mu}_n \cdot \hat{r}_{nm}
    \right) \left( \hat{\mu}_m \cdot \hat{r}_{nm} \right) \right],\\
    \label{eq:d1}
\end{equation}

$$  
  Q_{nm} = \frac{3\gamma}{2} \left[ \left( \frac{\sin (k_0 r_{nm})}{(k_0 r_{nm})} +
    \frac{\cos (k_0 r_{nm})}{(k_0 r_{nm})^2} - \frac{\sin (k_0 r_{nm})}{(k_0 r_{nm})^3} \right)
    \hat{\mu}_n \cdot \hat{\mu}_m +\right. \nonumber \\
    $$
\begin{equation}    
  -\left. \left( \frac{\sin (k_0 r_{nm})}{(k_0 r_{nm})} + 3\frac{\cos (k_0 r_{nm})}{(k_0 r_{nm})^2} -
    3\frac{\sin (k_0 r_{nm})}{(k_0 r_{nm})^3}\right) \left( \hat{\mu}_n \cdot \hat{r}_{nm}
    \right) \left( \hat{\mu}_m \cdot \hat{r}_{nm} \right) \right], 
    \label{eq:g1}
\end{equation}
where $\hat{\mu}_n :=  \vec{\mu}_n  /
\mu$ is the unit dipole moment of the $n$-th site and $\hat{r}_{nm} := \vec{r}_{nm}
/ r_{nm}$ is the unit vector joining the $n$-th and the $m$-th sites. 

Diagonalizing  the Hamiltonian (\ref{eq:ham}) we obtain the  complex eigenvalues $
\varepsilon_{n}=\mbox{E}_n-\mbox{i}\frac{\Gamma_{n}}{2}$
where $\Gamma_{n}$ is the radiative decay of the $n^{th}$ eigenstate. In general it differs from the radiative decay of the single molecule $\gamma$.   In particular, when the ratio $\Gamma_{n}/\gamma \gg 1$   we will talk about a ``superradiant state'' (SRS), otherwise when $\Gamma_n/\gamma \ll 1$ the  state is called ``subradiant''. In other words, a SRS  can radiate much faster than a single molecule, while a subradiant  one radiates at a rate much slower than the single molecule radiative decay. \\

Within the range of parameters considered here,   the imaginary part $Q_{ij}$ can be considered a small perturbation of the real part of the Hamiltonian (\ref{eq:ham}), moreover the system size is small compared to wavelength associated with the optical transition of the molecules (maximum size considered here is $L/\lambda_0 \approx 0.4$ ). 
In such case, the optical absorption of an eigenstate of the aggregate can be  estimated in terms of its dipole strength, computed only from the real part of the Hamiltonian (\ref{eq:ham}). 
Denoting the $n^{th}$ eigenstate of the real part of the Hamiltonian (\ref{eq:ham}) with $|E_n\rangle$, we can expand it on the site basis, so that 
\begin{equation} \label{eq:expan}
|E_{n}\rangle=\sum_{i=1}^{N} C_{ni} \, |i\rangle.
\end{equation}
Note that the site basis is referred to the molecules and is composed by the states $|i\rangle$, each of them carrying a dipole moment $\vec{\mu}_i$.  
If $N$ is the total number of molecules, then we will express the transition   dipole moment $\vec{D}_n$ associated with the $n^{th}$ eigenstate as follows: 
\begin{equation} \label{eq:dipst} 
\vec{D}_n=\sum_{i=1}^{N} C_{ni} \, \hat{\mu}_i. 
\end{equation} 
The dipole strength of the $n^{th}$ eigenstate is defined  by $|\vec{D}_n|^{2}$ (note that   due to normalization   $\sum_{n=1}^{N} |\vec{D}_n|^{2}=N$).  Under the approximation that the imaginary part of the Hamiltonian (\ref{eq:ham}) can be treated as a perturbation and $L/\lambda_0 \ll 1$ we have $|\vec{D}_n|^2 \approx \Gamma_n/\gamma$, which is valid for states with a large radiative decay rate (see  \ref{app-b} for a  comparison between dipole strengths and radiative decay widths for all models).

Thus, in the following we will consider only the real part of the Hamiltonian (\ref{eq:ham}):   
\begin{equation} \label{eq:hreal}
H_{r}=\sum_{i=1}^N e_0|i\rangle \langle i|+\sum_{i\neq j}\Delta_{ij}|i\rangle \langle j|.
\end{equation}
where $\Delta_{i,j}$  is given in equation~(\ref{eq:d1}).

 Finally we note that for small systems, when  $k_{0}r_{ij}\ll 1$, the Hamiltonian (\ref{eq:ham}) becomes 
\begin{equation}
\begin{array}{lll}
\label{real}
Q_{ij}&\simeq\displaystyle  \gamma \hat{\mu}_i \hat{\mu}_j,\\
&\\
\Delta_{ij} &\simeq\displaystyle \frac{\vec{\mu}_{i} \cdot \vec{\mu}_{j}-3(\vec{\mu}_{i} \cdot \hat{r}_{ij})(\vec{\mu}_{j} \cdot \hat{r}_{ij})}{r_{ij}^{3}}\\
\end{array}
\end{equation}
In this limit, the real term $\Delta_{ij}$  represents a dipole-dipole interaction energy with $\mu =|\vec{\mu}_j|$ and the  radiative decay $\gamma=\frac{4}{3}|\mu|^{2}k_{0}^{3}$. 
Nevertheless when the dimension of the aggregate becomes comparable with the wavelength $\lambda_0$ the dipole-dipole approximation fails. 
For the maximal sizes considered here($L/\lambda_0 \approx 0.4$) the dipole approximation can be considered good, even if 
there are already non-negligible deviations in some quantities between the dipole-dipole interaction equation~(\ref{real}) and the Hamiltionian in equation~(\ref{eq:hreal}), see~\ref{app-bc}.
For this reason in the following we will use  the expression given in equation~(\ref{eq:hreal}). 

\begin{figure}[t]
\includegraphics[scale=0.62]{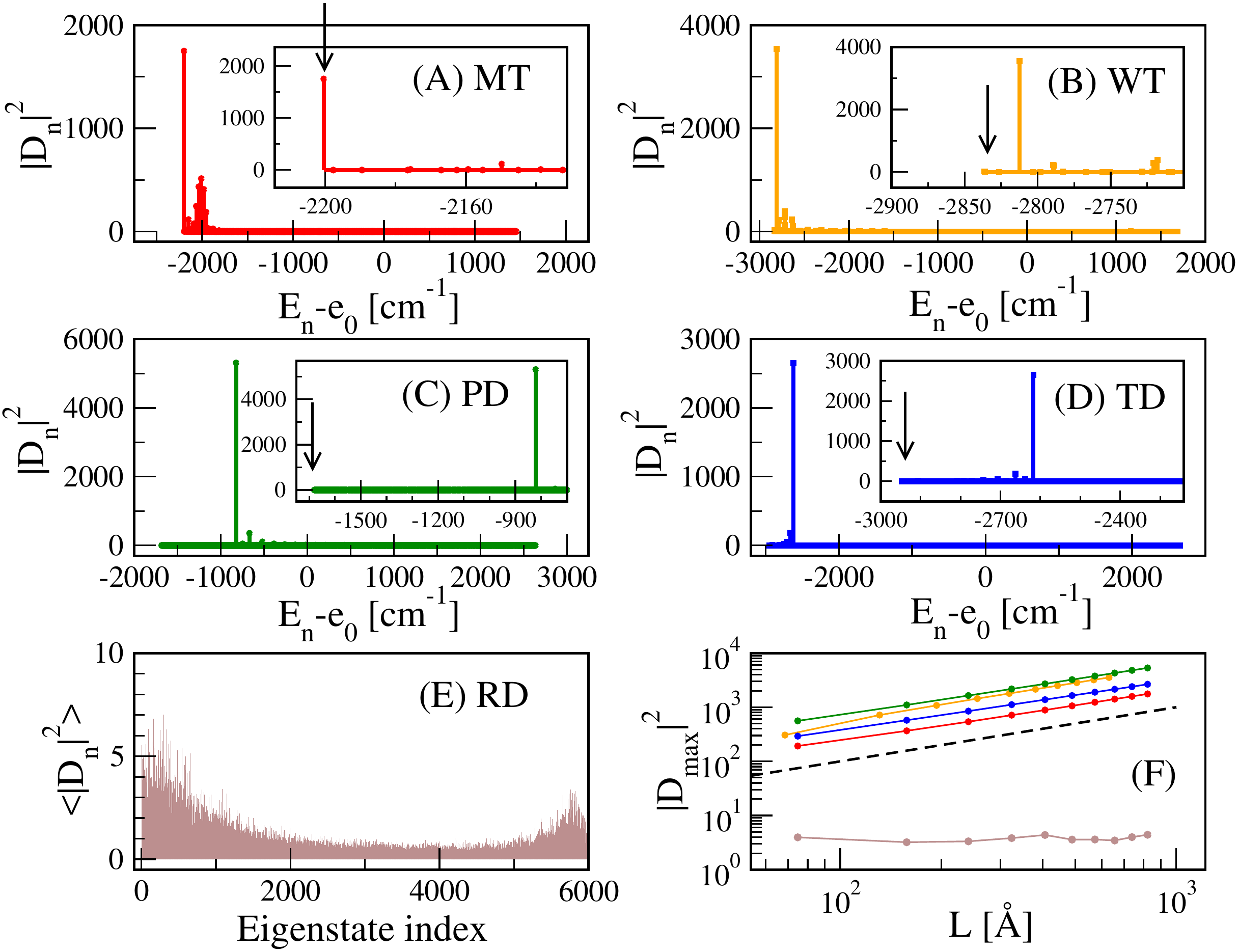}
\caption{(A) - (D) Squared dipole strength $|D_n|^2$ as a function of the energy $E_n-e_0$.  Superradiance arises in all cylindrical models since they are characterized by a high degree of geometrical symmetry. However, in the engineered structures made up of parallel and tangent dipoles (panels (C,D) ) the SRS does not coincide definitely with the ground state, nor it is close to it. On the other hand, in the MT model (A) the ground state is superradiant while in the WT model (B) the SRS, even if it does not coincide with  the ground state, it is indeed very close to it.
In panels (A,B,C,D) insets are shown with a magnification of the energy spectrum close to the SRS.
In the insets the arrows indicate the position of the ground state.
(E) Average squared dipole strength $\langle |D_n|^2 \rangle$ as a function of the eigenstate index.  The average  has been computed over 10 disorder realizations. 
(F) $|D_{max}|^2 $ as a function of the cylindrical length $L$. A linear dependence, as given by the dashed line $|D_{max}|^2 \propto L $, emerges clearly from all structures except the RD model (brown). Different colours stand for different models : MT (red), WT (orange), PD (green), TD (blue) and RD (brown).
In panels (A-E)  we considered cylindrical structures made of $6000$ dipoles. In panel (F) we considered cylindrical structures with a number of dipoles varying from $60$ to $6000$. 
  } \label{fig:secpar}
\end{figure}

\section{Single Cylindrical structures:  Results}
\label{sec:single}
In this Section we analyze first the collective dipole strengths of the eigenstates of the different models, showing the emergence of a superradiant state close to the ground state in natural complexes, see subsection~\ref{sub-41}. The coherence length is defined in subsection~\ref{sub-42} where also a new model with only nearest-neighbor couplings is introduced. Finally in subsection~\ref{sub-43} the results of our analysis about the thermal coherence length for the different models is shown. 
\subsection{Collective Dipole Strength}
\label{sub-41}
As a first goal let us analyze the dipole strengths associated with the eigenstates of the Hamiltonian models described in the previous section. 
%
For the five models introduced previously we diagonalized the  Hamiltonian in equation~(\ref{eq:hreal}), and  we analyzed in detail the dipole strengths $|D_n|^2$ of all eigenstates.  In Figure~\ref{fig:secpar}(A-E) we plot $|D_n|^2$ as a function of the energy E$_n-e_0$ of the corresponding eigenstate. 
All models but the random one (E) are characterized by the presence of  SRS in different positions  of the energy spectrum. For instance for the MT model the state having the largest dipole strength is the ground state while 
for the WT model it is very close to it. Note that the position of the superradiant state is below the excitation energy of a single molecule.  Since the dipole strength of the eigenstates determines the absorption spectrum~\cite{K31,K32}, a superradiant ground state implies a  red-shifted absorption spectrum which is a typical behaviour for molecular  J-aggregates~\cite{K31,K32,eisele,K2}. 
On the other hand for both the TD and PD models the SRSs are in the middle of the energy spectrum  (C,D).
Contrary to this general trend, the absence of ordering characterizing the random model (RD) 
does not guarantee the presence of SRS. 
Indeed it is well known that in the small volume limit $L/\lambda \ll 1$ symmetry is necessary to preserve super- and sub-radiance~\cite{haroche}. 

This is a clear indication that natural structures tend to push the SRS to the lowest energy region. Moreover, as the comparison with the other symmetric structure shows, this is not a trivial consequence of the symmetric arrangement. Other symmetric arrangements, such as the TD and PD, are still characterized by  SRS  but  ``living'' in an energy region far from the ground state.

SRSs are typically characterized by a collective dipole strength which grows with the length of the cylindrical structure.
This is  clearly shown  in   Figure~\ref{fig:secpar}(F)    where the maximal dipole strength $|D_{\rm max}|^2$
is shown as a function of the length $L$ of the cylinder.
As one can see the maximal dipole strength grows $\propto L$ for all models but the random one for  which it is  independent of $L$. \\

\subsection{Delocalized excitonic states at room temperature}
\label{sub-42}
Given a quantum state specified by the density matrix $\hat{\rho}$ it is possible to define its coherence length in the single excitation manifold defined by the  basis states $\ket{i}$~\cite{Schulten,Kosztin}: 
\begin{equation} \label{eq:lrho}
L_{\rho}=\frac{1}{N}\frac{\left(\sum_{ij}|\rho_{ij}|\right)^2}{\sum_{ij}|\rho_{ij}|^2}.
\end{equation}
The expression of   $L_{\rho}$  in equation~(\ref{eq:lrho})
measures how much a single excitation is spread coherently over the molecules composing the aggregate. 
To give an idea of its physical meaning let us consider three different simple cases: 
\begin{itemize}
\item   a pure localized state,  $\hat{\rho}=|i \rangle\langle i|$; then it is easy to see that the  coherence length defined in equation~(\ref{eq:lrho}) is given by $L_{\rho}=1/N$.
This case represents the minimal value that
$L_\rho$ can get.
\item   A completely delocalized mixed state  characterized by the density matrix: $\hat{\rho}=(1/N) \sum_{i=1}^{N} |i\rangle\langle i|$. In this case we have  $L_{\rho}=1$. This state is maximally delocalized in the basis, but it is completely incoherent. 
\item  Lastly we consider the fully delocalized  coherent state: $\hat{\rho}=(1/N) \sum_{i,j=1}^{N} |i\rangle\langle j|$.
In this case we have  $L_{\rho}=N$. Note that any pure state with constant amplitude $1/\sqrt{N}$ over the sites and arbitrary phases would give the same result.
\end{itemize}
Generally speaking we can see  that $1/N \leq L_{\rho} \leq N$.
The closer $L_{\rho}$ is to $N$, the higher a coherent delocalization  can be assigned to our state.
In the same way $L_\rho < 1 $ indicates an incoherent localized state. States characterized by $L_\rho \sim 1 $ have a little ambiguity (since both localization  and coherence are  measured on the same length scale). 

In what follows we will consider the previous models of cylindrical structures and we will compare them with an additional model where the positions of the molecule are the same of the MT model, but their interaction is only   nearest-neighbour. In this way we will be able to address the relevance of the range of the interaction to the thermal coherence length. 
For this purpose, let us consider a variant of the MT model, in which the Hamiltonian matrix elements are defined as follows:
\begin{equation}
\label{ham-nn}
H_{NN}=\left\{
\begin{array}{lr}
\sum_{i=1}^N e_0|i\rangle \langle i|+\sum_{i\neq j}\Delta_{ij}|i\rangle \langle j| & \mbox{ if } \quad r_{ij}\leq \bar{d}, \\ \\ 
0 & \mbox{ if } \quad r_{ij}>\bar{d}.
\end{array}
\right.
\end{equation}
where we have introduced the  cut-off distance $\bar{d}=9 \mbox{ \AA}$  and $\Delta_{i,j}$ is defined in equation~(\ref{eq:hreal}).   
In other words any lattice point interacts only with its four  nearest neighbours.

For all the models above  we have computed  the thermal coherence length at room temperature ($T=300K$), defined for a state at the canonical equilibrium and  whose matrix elements are given by:
\begin{equation} \label{eq:expand}
\rho_{ij}=\sum_{n} \frac{e^{-\beta E_n}}{\mbox{Tr}({e^{-\beta \hat{H}}})} \langle i|E_n\rangle \langle E_n|j\rangle,
\end{equation}
where $\beta=1/k_B T$. A very important question to be answered is how much the symmetrical arrangements that give rise to SRS
  are also able to produce a large thermal coherence length  at room temperature.

In that regard we calculate the coherence length $L_{\rho}$ according to   equation~(\ref{eq:lrho}), using a thermal density matrix equation~(\ref{eq:expand}),  as a function of the cylindrical length L for each of the cylindrical models studied so far, including the NN model described by equation~(\ref{ham-nn}).\\

As a final remark for this Section, let us  note that  for zero temperature
   $L_\rho$ depends only on how much the ground state is delocalized, while for infinite temperature we have a 
   fully mixed state  with: $\hat{\rho}=(1/N) \sum_{i=1}^{N} |i\rangle\langle i|$,
so that $L_{\rho}=1$ as explained above even if all eigenstates are fully delocalized.
  On the other hand at finite temperature the thermal coherence length is   determined by how much the energy eigenstates are delocalized on the site basis and also on how many eigenstates have an energy approximately within $k_B T$ above the ground state (i.e. from the density of states within an energy $k_B T$ from ground state).
For this reason, it is important to study the delocalization properties of the eigenstates of the nanostructures considered here. This analysis is shown in~\ref{app-pr}, where we show that the eigenstates of all models but the RD one have fully delocalized eigenstates with a very similar degree of delocalization.

\subsection{Thermal Coherence Length}
\label{sub-43}

It is usually thought that natural photosynthetic structures can support delocalised states even at room temperature because the nearest-neighbour (NN) coupling between the molecules is larger than the room temperature energy $k_B T \approx 200$ cm$^{-1}$. In Table~\ref{tab-02} we show the nearest-neighbour coupling for the different models considered here. As one can see these couplings are larger than $k_B T$, and the  maximal value between $\Omega_{1,2}$ are of the same order  amomg the different models. 
\begin{table}[t]
\centering
\begin{tabular}{|p{5cm}||c|c|c|}
\hline
& $ \Omega_1 $ [cm$^{-1}$] & $\Omega_2$ [cm$^{-1}$] \\
\hline
MT & 618 & 248 \\
\hline
WT & 115 & 629 \\
\hline
PD & 610 & 528  \\
\hline
TD & 1218 & 264 \\
\hline
\end{tabular}
\caption{Nearest-neighbour (NN) coupling for the different models. $\Omega_1$: azimuthal
 coupling for NN sites in the same ring (or between two adjacent
chains for the WT).
$\Omega_2$: vertical coupling for NN sites between rings (or in the same chain for the WT).}
\label{tab-02}
\end{table}

\begin{figure}[t]
\begin{center}
\includegraphics[scale=0.65]{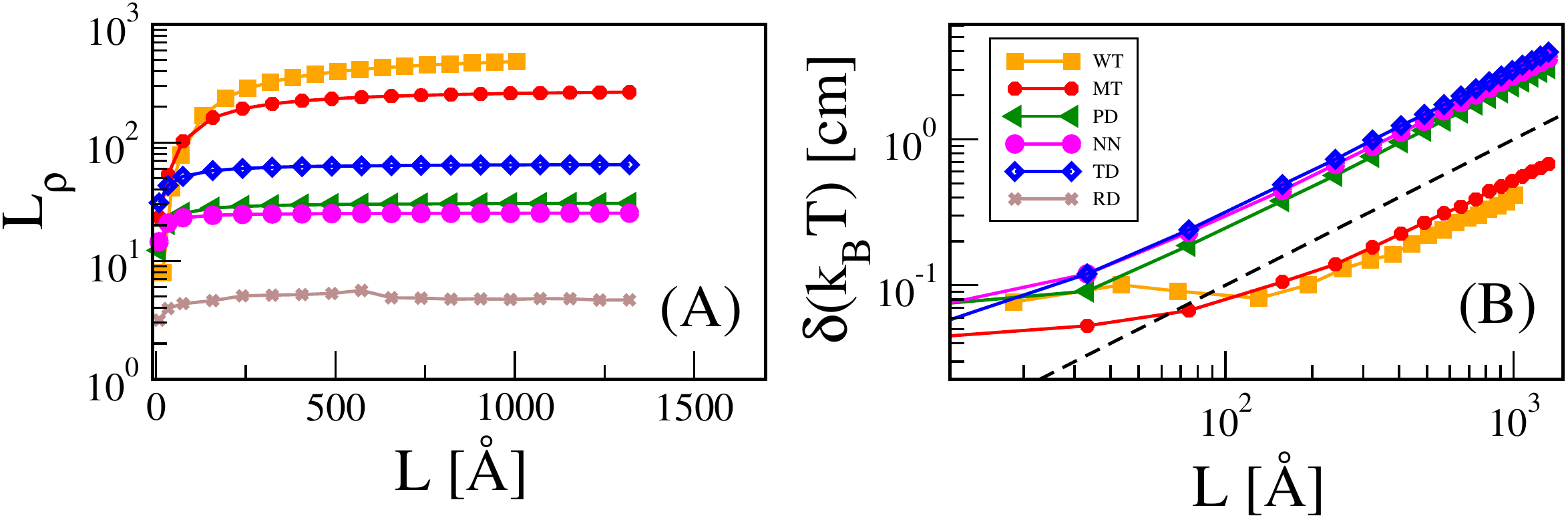} 
\caption{(A) Coherence length $L_{\rho}$ as a function of the cylindrical length in the 6 cylindrical models at $T=300$ K. The total number of chromophores $N$ varies from $120$ to $9600$.
(B) $\delta (k_B T)$, as given by equation~(\ref{deltae-eq}), as a function of the cylindrical length $L$ at a fixed temperature $T=300$ K. All models have a total number of dipoles ranging from $120$ to $9600$. Note that since energy is measured in $[\mbox{cm}]^{-1}$,
the mean energy density in the thermal energy width $k_B T$ is measured in $[\mbox{cm}]$, see equation~(\ref{deltae-eq}). 
} \label{fig:lrho}
\end{center}
\end{figure}

Let us now consider the thermal coherence length of the structures analyzed here at room temperature. 
Figure~\ref{fig:lrho}(A) shows the dependence of $L_{\rho}$ on the cylinder length $L$  (with a corresponding  number of dipoles $N$ ranging from $120$ to $9600$). \\ In all models but the RD, the coherence length $L_{\rho}$ increases quite markedly for small $L$   until it   reaches a plateau for larger $L$ values. Apart from the RD structure,   that exhibits a coherence length $L_{\rho}\approx 1$, the other structures are characterized by  $1 \le L_{\rho} \le \mbox{N}$. 
This means that the thermal state at room temperature of these structures has  a high degree of excitonic delocalization.
Moreover  it emerges clearly that the natural complexes (MT and WT) show the highest values of thermal coherence length if compared with  the other engineered structures.
It is interesting to note that the MT complex supports a coherent delocalisation of the excitation over hundreds of molecules even at room temperature, which is one order of magnitude larger than the delocalisation supported by  the NN
model despite the fact that in the NN model the molecules  have the same position and the same nearest-neighbour  coupling of the MT model. This shows that the ability of such structures to support large delocalised excitation even at room temperature goes beyond the strength of the NN coupling between their molecules.  From Figure~\ref{fig:lrho}(A) we can also deduce that the large coherence length of the natural systems cannot be explained by the presence of long range interactions. Indeed long-range interactions are present also in the PD and TD models, but their thermal coherence length is one order of magnitude smaller. 
By comparing the different cylindrical  structures, 
one may also observe that the further the SRS is from the ground state, the lower is $L_{\rho}$. 
One could argue that natural structures concentrate the most radiative states (states with the largest dipole strength) close to the ground state in order to maximize their thermal coherence length. We will discuss the relationship between the presence of the SRS close to the ground state and a large coherence length in the next Section.


The presence of a large thermal coherence length  can be related to the structural properties of the energy spectrum. To this end we consider the mean energy density $\delta(k_B T)$ defined as the number of states contained in a unit of thermal energy $k_B T$, i.e.
\begin{equation}
\label{deltae-eq}
  \delta (k_B T) = \frac{1}{k_B T} \int_{E_1}^{E_1 + k_B T} \, N(E) \, dE,
\end{equation}
where $E_1$ is the ground state and $N(E)$ is the density of states (number of states per unit energy). 
In particular, we would like to study the dependence of the average density of states, equation~(\ref{deltae-eq}) on the  cylindrical length $L$.
Results are shown in Figure~\ref{fig:lrho}(B) and clearly indicate that, not only, in general, the  average density increases proportionally to $L$, but more important,   natural structures are characterized by the smallest average densities
(approximately one order of magnitude less than the other structures).
Such a low density of states in the lower part of the spectrum induces, see Figure~\ref{fig:lrho}(B), an enhanced thermal coherence length. Indeed, if all the eigenstates have approximately the same degree of delocalization, as measured by their PR for instance, then for a smaller  number of states within an energy $k_BT$ from the ground state, the thermal coherence length is larger, as explained above.
In order to explain the origin of the low density of states, let us observe that  : ($i$) it cannot be due to the intensity of the NN coupling. Indeed the NN model, which has the same NN coupling as the MT model, has a much higher density of states and a smaller thermal coherence length; ($ii$) it cannot be due to the range of interaction  since also the TD and PD model  are characterized by the same  interaction range  but they display a higher density of states and as a consequence a smaller thermal coherence length.  
Below we propose an explanation of the connection between the presence of a SRS close to the ground state and a low density of states, implying a large thermal coherence length. 

\section{Relationship between structure and macroscopic coherence}
\label{sec:rel}

In this section  we propose an  explanation of why such a low density of states is connected to the presence of  SRS close to the ground state of the system. 
As we will show below the low energy part of the spectrum for both  the MT and WT models arises from a super-transfer coupling between states with a large (giant) dipole belonging to some sub-unit of the whole cylinder. In the case of MT we will show that the super-transfer coupling arises between giant dipole eigenstates of single rings, while in the case of WT   the super-transfer arises between eigenstates belonging to different sub-units of the whole cylinder.  The presence of super-transfer induces a large coupling energy  which  decreases the  density of states. 
As a  clear signature of this, we  show below that super-transfer is also able to induce the  emergence of an energy gap between the ground state and the first excited state. 

Specifically in subsection~\ref{sub-51} we analyze cylinders made of a sequence of rings and we show that the symmetry present in the system implies that each eigenstate of a ring couples only to a correspondent eigenstate of the other rings. We also show that the dipole strength of the eigenstates of each ring is concentrated in few superradiant states.
In subsection~\ref{sub-52} we show that the coupling between superradiant states in each ring displays a super-transfer effect, while the coupling between the subradiant states   is characterized by a sub-transfer effect. Finally in subsection~\ref{sub-53} we show how in natural structures the super-transfer coupling produces a depressed density of states close to the ground state, thus enhancing the thermal coherence length.

\subsection{Structure of ring eigenstates coupling} 
\label{sub-51}

\begin{figure}[t]
\begin{center}
\includegraphics[scale=0.65]{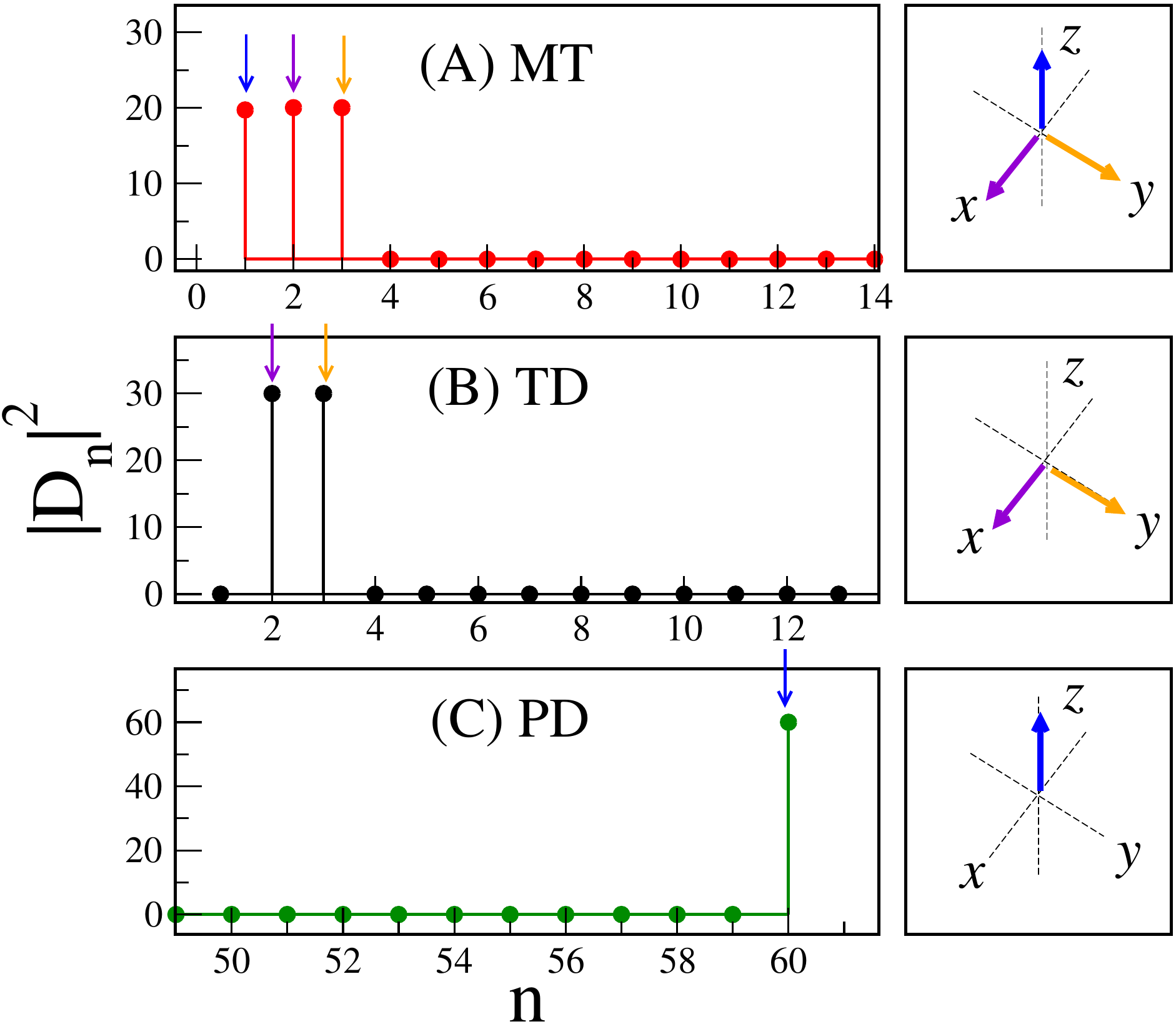}
\caption{Dipole strength of few eigenstates (in the lowest or highest part of the energy spectrum) {\it vs.} the eigenstate index $n$, for a single ring composing three different nanotubular structures: MT, TD and PD. Lateral panels indicate the spacial direction of the giant dipoles of the SRS.  Each ring of the  three structures considered (A,B,C) is composed by $N_2=60$ dipoles.} \label{fig:sr}
\end{center}
\end{figure}

In order to analyze the super-transfer effect, let us consider the properties of the eigenstates of the single rings composing three different nanotubes:  MT, TD and PD. 
All the above mentioned models are composed of a sequence of rings, each containing 60 molecules, as explained in Section~\ref{sec:mod}. The case of the WT model will be discussed later since its structure is more complicated. 
In Figure~\ref{fig:sr} the dipole strength of few eigenstates (ordered from low to high energy) of a single ring, containing 60 dipoles, is shown for the different structures. Note that the sum of all the dipole strengths must be equal to the number of the dipoles in the ring $N_2=60$ as explained in the previous Sections. As one can see in the  MT case the whole dipole strength is concentrated in the lowest three eigenstates, each having a dipole strength approximately equal to $N_2/3$. Each dipole strength is oriented in a different spatial position with the ground state having a dipole strength along $z$ corresponding to the direction of the cylinder axis and the other two states perpendicular to it in the ring plane, see inset in Figure~\ref{fig:sr}(A) . 
In the TD model in Figure~\ref{fig:sr}(B),  the dipole strengths are concentrated in the first and second excited state (which are degenerate and having $|D_n|^2=N_2/2$ each) and their direction lies in the plane perpendicular to the direction of the cylinder axis.  Finally for the PD model in Figure~\ref{fig:sr}(C),    the whole dipole strength is concentrated in the most excited state and it is directed along the \textit{z} axis (cylinder axis). 

A common feature of these structures is their invariance under a $2 \pi /N_2$ rotation around the cylinder axis. Strictly speaking, in the MT model such symmetry is slightly broken due to the presence of alternating $ \alpha $ angles, see~\ref{app-a}.  Nevertheless since $\alpha$ is very small the change due to the symmetry breaking is negligible.
As a consequence the Hamiltonian for each ring is a circulant matrix,  i.e. each row can be obtained by a cyclic permutation of the previous one. Circulant matrices are diagonalized by the Fourier basis, so that  the components of the eigenstates of each ring $|\varphi_q\rangle$   on the site basis $|j \rangle$ are given by 
\begin{equation}
\langle j |\varphi_q \rangle = \frac{1}{\sqrt{N_2}} e^{i 2 \pi jq/N_2} \quad {\rm for} \quad q=1,...,N_2.
\label{eq:ring}
\end{equation}
 Due to the rotational invariance   the coupling matrix between two rings is also circulant.
 
\begin{figure}[t]
\begin{center}
\begin{tabular}{cc}
    {\bf \Large(A)} & {\bf \Large(B)} \\
    \includegraphics[scale=0.39]{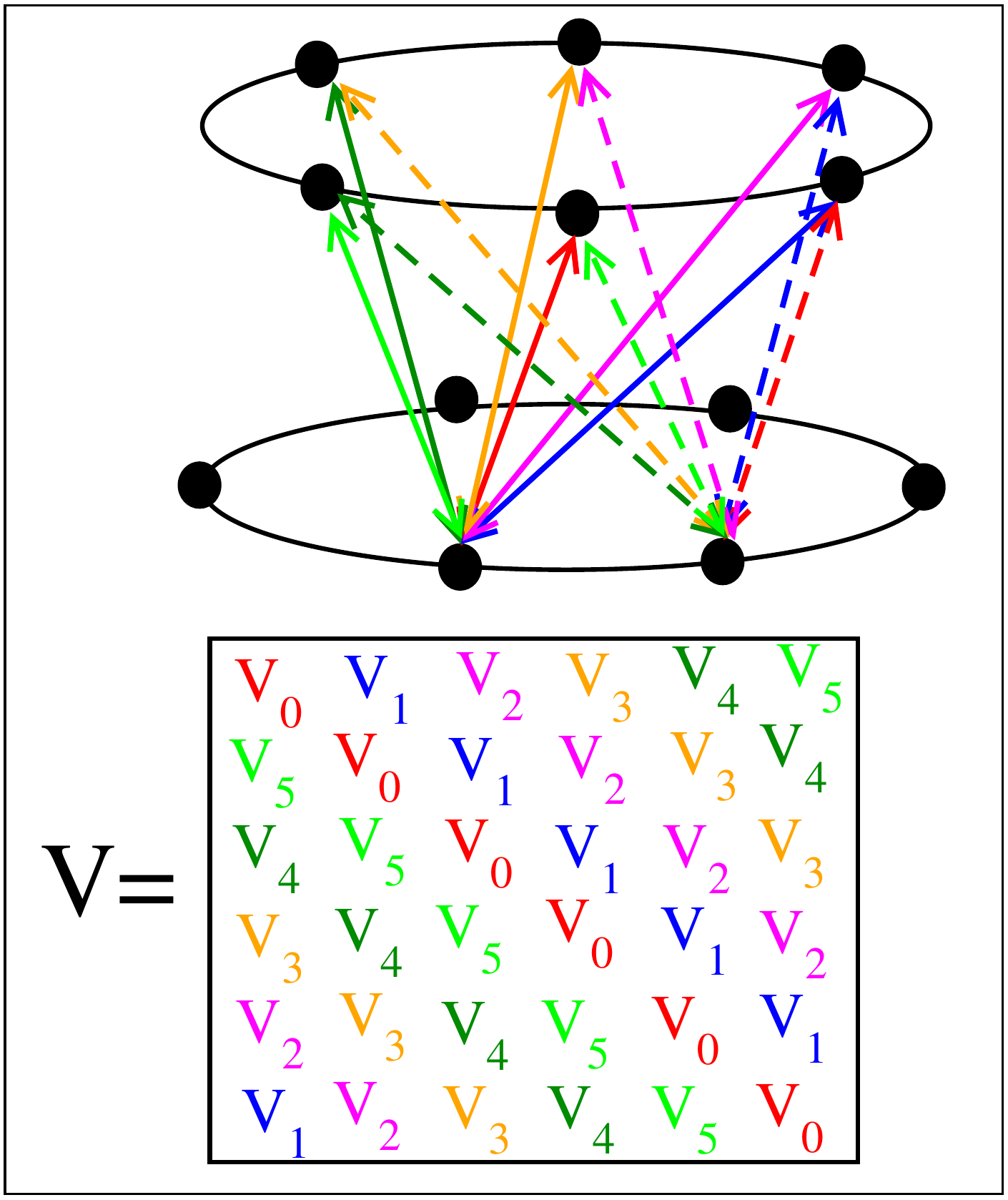}
 & \includegraphics[scale=0.39]{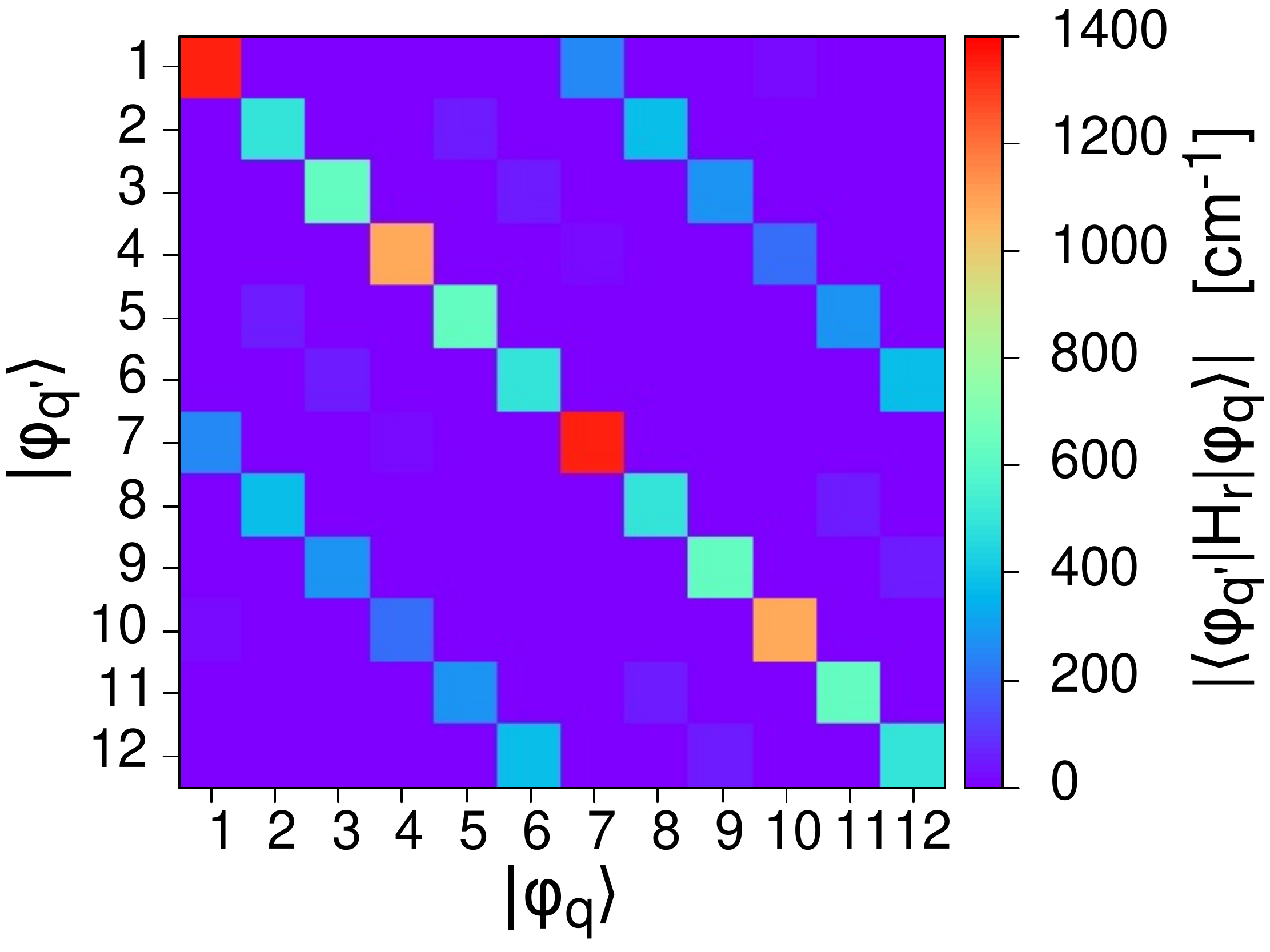}
 \\
\end{tabular}
\caption{(A) Graphical representation of the  coupling between the sites of two rings, each formed by six molecules. Same colours indicate the same couplings. The circulant coupling matrix $V$, see equation~(\ref{hblock}), generated by the symmetric coupling is represented below. (B) Modulus  of the Hamiltonian $H_r$~(\ref{hblock}) matrix elements for the MT model, for the case sketched in (A) in the Fourier basis. Each ring eigenstate is mainly coupled only to one corresponding eigenstate in all the other rings. 
} \label{fig:MT3}
\end{center}
\end{figure}

To make explicit this point, let us work out a specific example of 
  two rings.  The Hamiltonian reads: 
\begin{equation}\label{hblock}
   H_r=
  \left[ {\begin{array}{cc}
   D & V \\
   V & D \\
  \end{array} } \right]
\end{equation}
where $D$ refers to the Hamiltonian   of a single ring (which is diagonal in the Fourier basis given in  equation~(\ref{eq:ring}))  and $V$ represents the interaction  between two rings.
 The total Hamiltonian matrix $H_r$  can be made block diagonal  by the  matrix
 
\[
   U_r=
  \left[ {\begin{array}{cc}
   U & 0 \\
   0 & U \\
  \end{array} } \right]
\]
where the  elements of $U$ are given by equation~(\ref{eq:ring}): $ U_{j,q} = \langle j |\varphi_q \rangle $.
In other words,
  each ring eigenstate   is coupled only with one corresponding eigenstate of any  other ring.  This is clearly shown in Figure~\ref{fig:MT3}(B), where the  matrix elements of the  Hamiltonian of a small cylinder composed of two rings of 6 sites each, are represented in the basis given by the tensor product of the Fourier basis of each ring. As one can see, this results in a block structure where each block has only diagonal elements.
 
\begin{figure}[t]
\begin{center}
\includegraphics[scale=0.36]{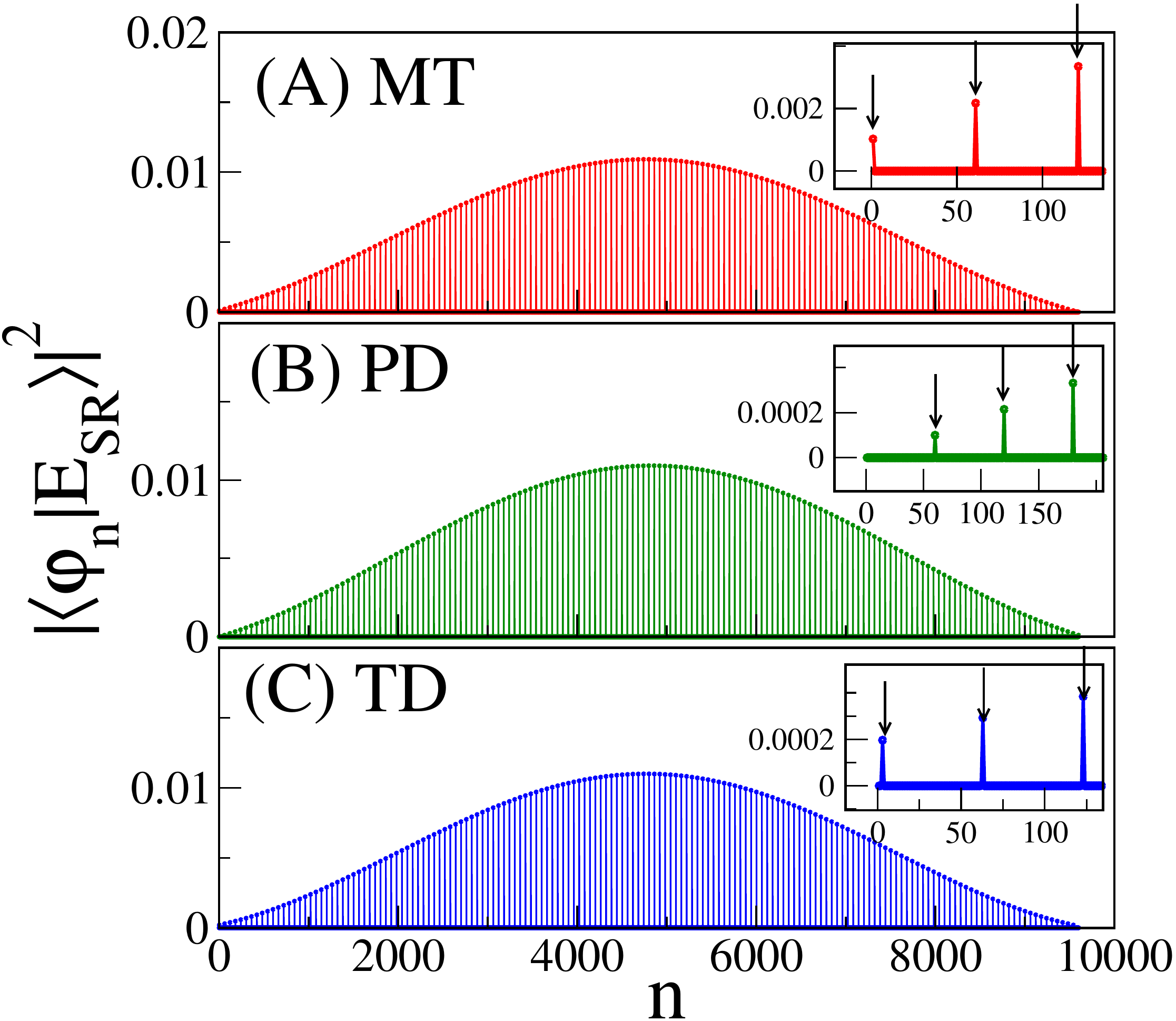}
\includegraphics[scale=0.45]{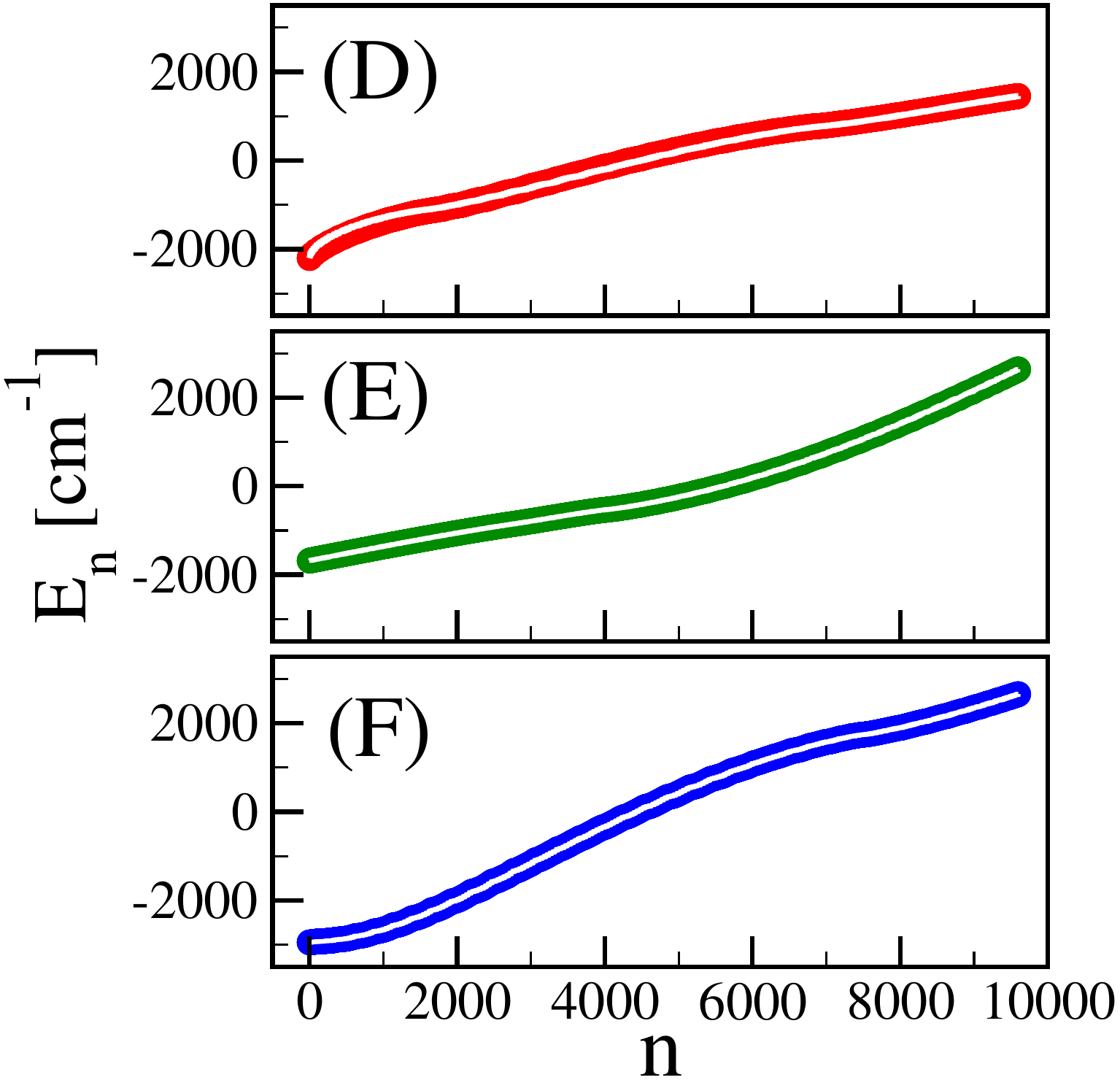}
\caption{Left Panels: Projections of the most SRS  of the whole cylinder $|E_{\rm SR}\rangle$
over the single ring eigenstates $|\varphi_n\rangle$ as a function of the eigenstate index n. In each case we selected a total number of dipoles $N=9600$ (then n$=1,...,9600$), which corresponds to $N_1=160$ rings and $N_2=60$ molecules in each ring. 
(A) MT model, (B) PD model, (C) TD model (in the insets  the corresponding  blow up of the low energy part of the energy spectrum). Arrows refer to the SRS of the single rings. 
Right Panels: energy
spectrum of the three different cylindrical structures: 
(D) MT model, (E) PD model, (F) TD model.
 Coloured symbols represent the  exact numerical spectrum, white lines stand for the spectrum obtained from  the  analytic approximate eigenstates, see equation~(\ref{tenseig}). 
} \label{fig:fb}
\end{center}
\end{figure}

As a consequence of the symmetric structure of the nanotubes considered above,  all the eigenstates of the whole cylinder can be ``generated'' by the coupling between the eigenstates of single rings, see also discussion in \cite{K5}. Specifically the SRS of the whole cylinder is generated by the coupling of the SRS of the single rings. 
In order to prove  that,  we  show  in  Figure~\ref{fig:fb}(A, B, C)  the most SRS for  the  different models projected along  the eigenstates of the single rings. In the figure we considered cylinders made of $N_1=160$ rings, with $N_2=60$ molecules per ring, for a   total number of dipoles of $N=9600$.
Let us analyze the single models individually:
\begin{enumerate}
\item For the  MT model, one can see that  the most SRS (having a dipole along the cylinder axis) has components only on the ground states of the single rings (indicated by arrows in the inset of Figure~\ref{fig:fb}(A)) that are also
  SRS  with a dipole strength  along the $z-$axis, see Figure~\ref{fig:sr}(A).
\item In the PD model, Figure~\ref{fig:fb}(B), the most SRS, $|E_{2814}\rangle$, projects itself on the most excited state in the single ring spectrum, which corresponds to the only SRS  of the PD ring, see Figure~\ref{fig:sr}(C). Note that  $|E_{2814}\rangle$ indicates the $2813^{rd}$ excited state.
\item In the TD model  there are two most SRS which are degenerate with a different polarization: one along the $x$ direction and one along the $y$ direction. In Figure~\ref{fig:fb}(B) we considered only the SRS with a polarization along the $y$ direction, which corresponds to the  state  $|E_{1083}\rangle$. Such state  has non zero projections only onto  the  second excited states of the single ring with the same dipole direction of the SRS of the whole cylinder, see Figure~\ref{fig:sr}(B). Correspondingly the other SRS with a polarization along the $x$ direction will have projection only on the SRS of the single ring with the same polarization. 

\end{enumerate}

These findings allow for a further approximate scheme for the eigenstates of the cylindrical structures considered above. 
Indeed, since each eigenstate of any single ring is coupled only to a corresponding eigenstate of the other rings,
  we can decompose the whole cylinder into independent chains where each site of the chain corresponds to a single ring eigenstate.
  For a chain having $N_s$ sites and  nearest-neighbour interactions the eigenstates are independent of the coupling and   given by:   
 
\begin{equation}
\label{eq:nn}
\langle k | \psi_r \rangle = \sqrt{\frac{2}{N_s+1}} \sin \left( \frac{\pi k r}{N_s+1} \right),
\end{equation}
where $k$ represents the site index and $r=1,..,N_s$.  Clearly when the interaction range is not  nearest-neighbour, the above expression for
 the eigenstates is no longer valid.   Nevertheless for the natural structures considered in this paper the interaction 
  is short-range, decaying as $1/r^3$ for the realistic cylinder length considered here, so that in a first scheme we can consider the nearest-neighbour eigenstates as a good approximation.  
  Note however that care should be taken to generalize such approximation since the interaction between the molecules is much more complicated than a simple dipole-dipole one. 
  For instance the coupling is also affected by the dipole strength of the ring eigenstates involved as we will see below. Nevertheless we can assume that the chain of eigenstates is diagonalized by the  same eigenstates of a chain with nearest-neighbour coupling  for the parameters and the realistic system sizes considered here. 
  Building the eigenstates as a tensor product between the Fourier basis for the ring~(\ref{eq:ring}) and the one for the chain~(\ref{eq:nn})
  \begin{equation}\label{tenseig}
      \braket{j,s}{\Psi_{q,r}} = \frac{1}{\sqrt{N_2}} e^{i 2 \pi jq/N_2} \sqrt{\frac{2}{N_1+1}} \sin \left( \frac{\pi s r}{N_1+1} \right)
  \end{equation}
  (with $j,q=1,\dots,N_2$ and $s,r=1,\dots,N_1$) we can diagonalize the Hamiltonian 
  of the whole cylinder in order to obtain an approximation of the actual spectrum of the whole structures. The results are
   shown in Figure~\ref{fig:fb}(D, E, F) where  the spectrum obtained from  exact numerical diagonalization is compared with the spectrum obtained by diagonalizing the matrix   
  with the  eigenbase in equation~(\ref{tenseig}). 
 As one can see,  the proposed  analytic basis gives an excellent approximation
  of the spectrum obtained by exact numerical diagonalization. 

\subsection{Super and Sub-Transfer} 
\label{sub-52}

In the previous section we have shown that each eigenstate of a single ring couples only with a corresponding eigenstate of the other rings (apart for a small symmetry breaking factor  present in the MT model). Here we will show that the coupling between the eigenstates with a large dipole strength is enhanced with respect to the coupling between the single molecules within each ring by a factor proportional to the number of molecules placed on each ring. Such effect is known  in literature as super-transfer~\cite{srlloyd}. 
At the same time we will show that the coupling between the eigenstates of the single rings with a small dipole strength is suppressed with respect to the coupling between the single molecules, giving rise to another collective sub-transfer effect, which has not been fully addressed in literature.

In order to prove the previous statements,  let us compute the coupling strength between two eigenstates of two rings, say 1 and 2.
Let us indicate the two corresponding $q$-th eigenstates of the two rings as $$|\psi^{s,q} \rangle= \sum_k C_k^{s,q} |k\rangle,$$ where the states $|k \rangle$ represent  the site basis of a ring  and $s=1,2$.  The coupling between two single ring eigenstates belonging to two different rings can be written as:
\begin{equation}
\label{eq:st}
V^q_{12} = \langle \psi^{1,q}| V| \psi^{2,q } \rangle= \sum_{k,k'} (C^{1,q}_k)^* C^{2,q}_{k'} V_{k,k'}.
\end{equation}
Using  equations ~(\ref{eq:hreal}) we have  that  $V_{k,k'}=\Delta_{k,k'}= f(r_{k,k'}) \vec{\mu}_k \cdot \vec{\mu}_{k'} + g(r_{k,k'}) (\vec{\mu}_k \cdot \hat{r}_{k,k'}) (\vec{\mu}_{k'} \cdot \hat{r}_{k,k'})$. When the distance between the two rings is much larger than their diameter we can approximate $r_{k,k'} \approx R_{12}$ where $R_{12}$ is the distance between the centres of the two rings. In this limit,  equation~(\ref{eq:st}) becomes
\begin{equation}
V^q_{12} =  \sum_{k,k'} (C^{1,q}_k)^* C^{2,q}_{k'} \left[ f(R_{12})\vec{\mu}_k \cdot\vec{\mu}_{k'} + g(R_{12})   (\vec{\mu}_k\cdot \hat{R}_{12}) (\vec{\mu}_{k'}\cdot \hat{R}_{12})\right] \, ,
\end{equation}
which can be expressed in terms of the dipole strengths using equation~(\ref{eq:dipst})
\begin{equation}
\label{eq:st2}
V^q_{12} =  \left[ f(R_{12}) |\vec{D_q}|^2  + g(R_{12})   (\vec{D}_q\cdot \hat{R}_{12}) (\vec{D}^*_{q}\cdot \hat{R}_{12})\right] \, .
\end{equation}

\begin{figure}[t]
\begin{center}
\includegraphics[scale=0.6]{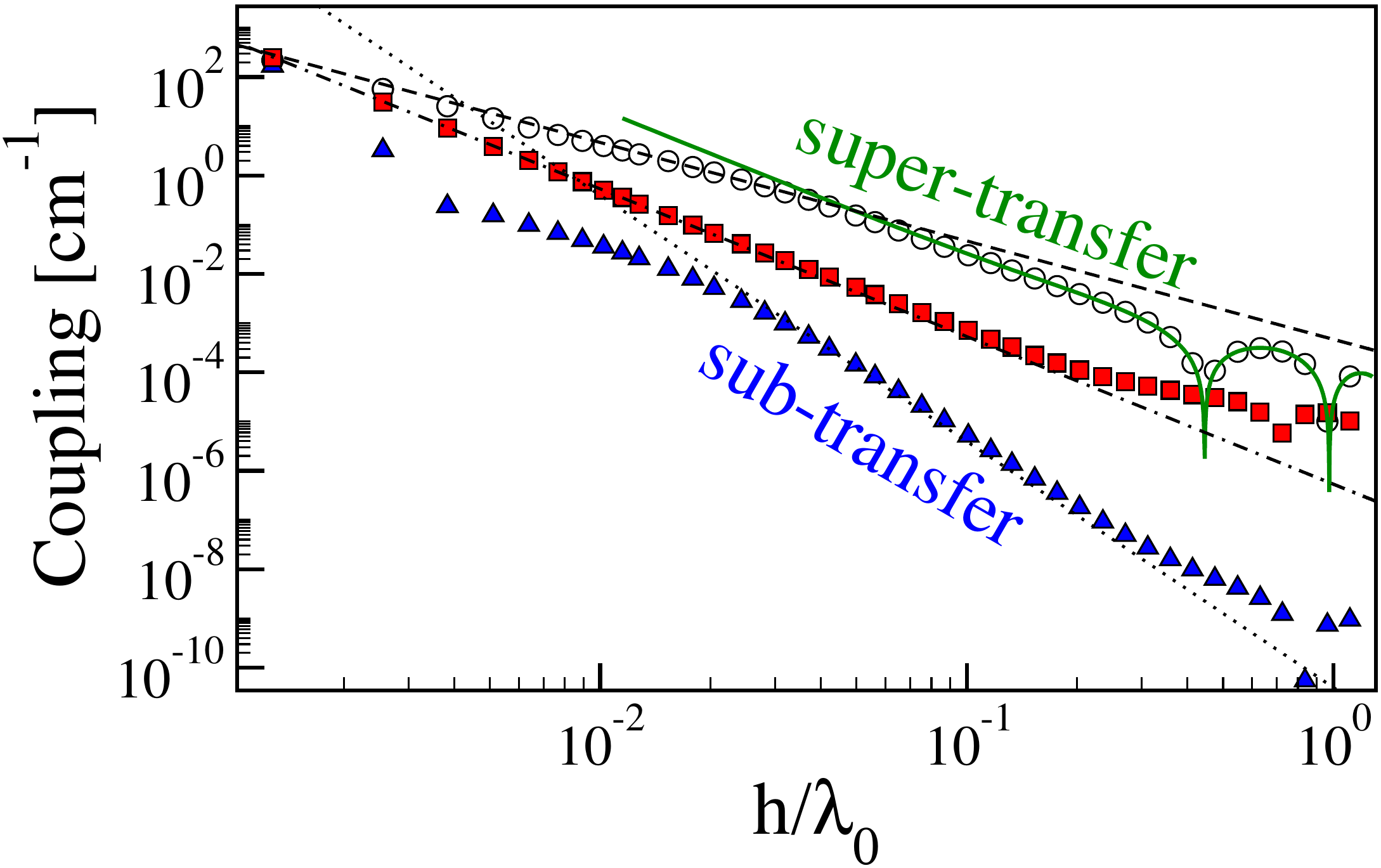}
\caption{Coupling between ring eigenstates  as a function of their distance $h$ normalized to the wavelength $\lambda_0 =650$ nm.
Open circles represent the coupling $V^q_{12}$ (see equation~(\ref{eq:st})) between  the ground states of two rings for the MT model.
Red squares stand for the maximal coupling between individual molecules in the two rings.
Blue triangles represent the coupling between the most excited eigenstates of the two rings.
Green curve represents the coupling between the giant dipoles of the ground states as given by  equation~(\ref{eq:st2}).
The three lines represent respectively the behaviours $1/r^2$ (dashed), $1/r^3$ (dot-dashed), $1/r^5$ (dotted). 
} \label{fig:sr2}
\end{center}
\end{figure}
 As a result, we obtain    $ V^q_{12} \propto |D_q|^2 \propto N_2$. In other words the  eigenstates with a large dipole strength will have a coupling enhanced by a factor proportional to the number of molecules $N_2$ in the ring.  
 
The above expression   represents the interaction between the giant dipoles of the eigenstates of each ring. Therefore states with a large dipole strength will have a super-transfer coupling, (proportional to the dipole strength of the eigenstates)   increasing linearly with the number of molecules $N_2$ in each ring. At the same time, the coupling between two eigenstates with zero dipole strengths will be  suppressed, leaving only higher order multipole terms to contribute to the coupling. This will lead to a sub-transfer coupling. 
The super and sub-transfer effects for the MT model are shown in  Figure~\ref{fig:sr2} where we compare: $(i)$ the coupling between the superradiant ground states (which have a large dipole strength) of two rings as a function of their rescaled distance (open circles); $(ii)$ the maximal coupling between single molecules of each ring as a function of the distance between the two rings (red squares); $(iii)$ the coupling between the most excited states (with a very small dipole strength) of each ring as a function of their distance (blue triangles).

Let us  comment in detail this figure. First of all we note that 
the coupling between the states with a large dipole is clearly larger (by a factor $\sim N_2 = 60$) than the maximal coupling between the single molecules thus showing the super-transfer effect.
Moreover, the coupling between the eigenstates with a small dipole strength is much  smaller than the maximal coupling between single molecules: this shows the   sub-transfer effect.

In the same figure, as a continuous green curve we show   the coupling between the ground states as given  by equation~(\ref{eq:st2}). As one can see, at sufficiently large distance,  the couplings are well approximated by  equation~(\ref{eq:st2}) thus  confirming that the coupling is  enhanced by a factor proportional to the number of molecules in each  ring $N_2$. 

Another important observation concerns the dependence of such couplings from the distance $r=h/\lambda_0$ and how it is modified
by   the super and sub-transfer effect. 
We can distinguish three different regimes: at small distances, at intermediate distances and at distances comparable with the wavelength of the optical transition. At large distances, when $h \sim \lambda_0$  an  oscillatory behaviour arises due to the presence of  oscillatory terms in the Hamiltonian of the system, see equation~(\ref{eq:hreal}). At intermediate distances the super-transfer coupling decays with $1/r^{3}$ as the coupling between single molecules, consistently with the dipole-dipole nature of the interaction.
On the other hand, the sub-transfer coupling decays as $1/r^5$ which is consistent with high order multipole expansion of the  coupling since the dipole interaction is suppressed. 
At small distances the behaviour of the coupling with distance is less trivial: while the single molecule coupling still behaves as $1/r^3$, the sub-transfer coupling decays much faster and then it goes as $1/r^5$ as explained above. On the other hand the super-transfer coupling decays as $1/r^2$, which is  much slower than the dipole coupling. Since all the couplings start from the same intensity at very small distances and the superradiant one has to go above the single molecule coupling, it makes sense that its decay is slower than $1/r^3$, but further analysis is needed to understand the origin of such slow decay of the interaction between giant dipoles.  

\subsection{Super-Transfer and density of states} 
\label{sub-53}

\begin{figure}[t]
\begin{center}
\includegraphics[scale=0.6]{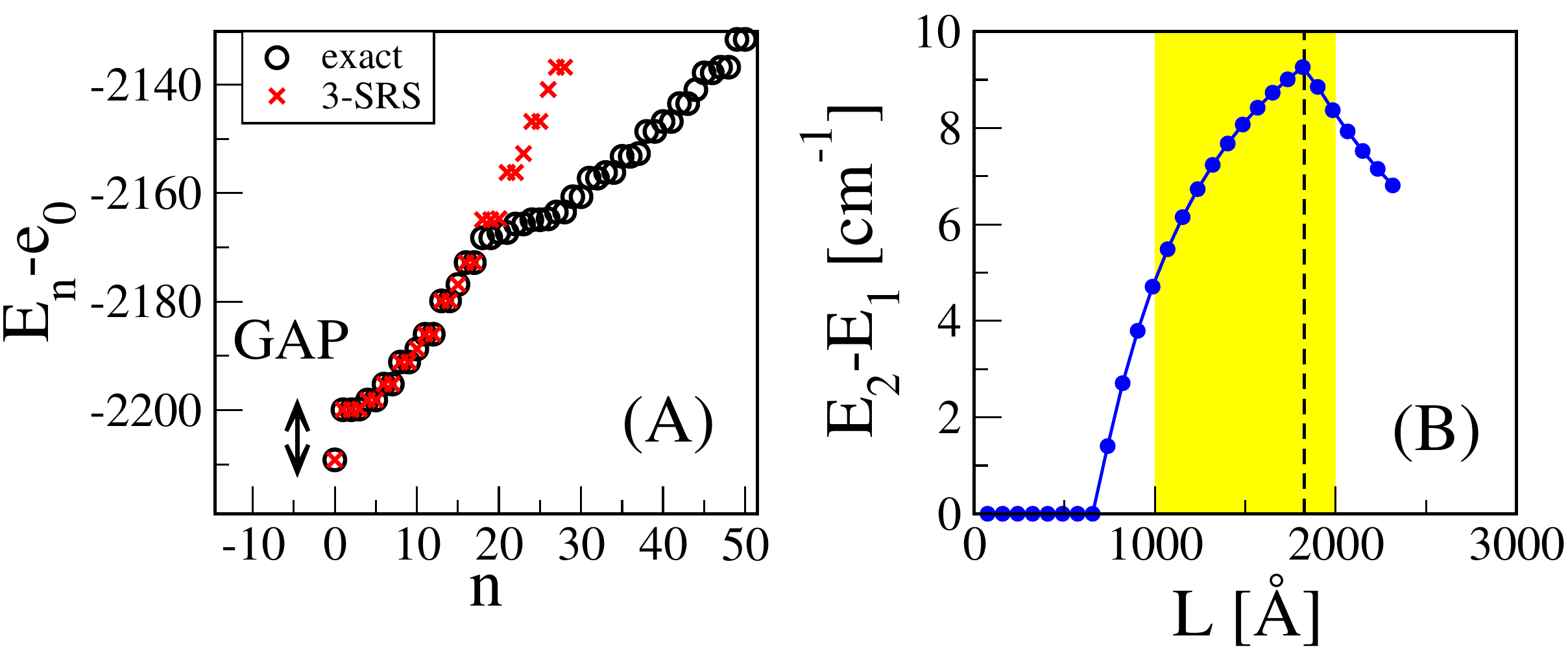}
\caption{(A) The lowest part of the energy spectrum for a MT nanotube with  220 rings (open circles)   compared with the 
spectrum generated by the super-transfer coupling between the the three most SRSs  of each ring (crosses). 
 Note   the presence of a consistent energy gap between the ground state and the first excited state. (B) Energy gap (distance between the ground   and the first excited state)  for the MT model   as a function of the nanotubular length. As one can see there is a region where the gap increases with the system size. Maximal gaps occurs at $L =  1826 \, \mbox{\AA}$. The yellow vertical strip  indicates the region where natural complexes operate. } 
\label{fig:gap}
\end{center}
\end{figure}

From the discussion above we can conclude that all the SRS belonging to each ring will couple between themselves through a
super-transfer coupling. 
For instance, in the case of the MT model, also the other two SRS of the single rings corresponding to the first and second excited states will couple between themselves by super-transfer, see Figure~\ref{fig:sr}(A). While for the PD and the TD model the coupling between the SRS of the rings gives rise to the SRS of the whole cylinder which lies far away from the ground state, for the MT model the coupling between the SRS of the single ring determines completely the lowest part of the spectrum.
In order to prove the last sentence we consider the $3 N_1$ eigenvalues generated by  the super-transfer coupling of the  three SRS for each ring of the MT model. The spectrum generated by the three SRS is shown
in  Figure~\ref{fig:gap}(A) together with the exact spectrum of the MT model. As one can see this simple approximation allows to compute
with high accuracy the lowest energy part of the spectrum. The presence of super-transfer induces a large coupling energy in the lowest part of the spectrum, which in turn diminishes the density of states. This is also signaled in  Figure~\ref{fig:gap}(A) by the change of slope seen in the lower part of the spectrum.
A further evidence of such decreased density of states induced by the super-transfer coupling of the SRS of each ring is
shown in Figure~\ref{fig:gap}(B). Here  the energy gap between the ground state and the first excited state for the MT model
is shown  as a function of the length of the nanotube. Contrary to what can be expected for generic systems, the energy gap
   increases with the system size instead of decreasing,   up to a critical system size, above which it decreases.
The  maximal energy gap occurs at a distance of $\sim 182.6 $ nm which is compatible with the typical length of such nanostructures found in nature, ranging  between $100$ and $200$ nm. Note that it would be interesting to understand the critical system size at which the gap has a maximum. We intend to study this problem in a future work. 

The results obtained so far can be generalized to more complicated structures, such as the WT model, as the preliminary results shown in \ref{app-wt} show. 
Indeed even for the WT model, where the disposition of dipoles is much more complicated than in the previous models, one can show that the superradiant state close to the ground state emerges from the supertransfer coupling between the superradiant states of cylindrical sub-units of the whole cylinder.

Summarizing,
the   analysis both for the MT and  the WT models  show how a precise ordering of the dipoles in these systems can favour the emergence of super-transfer between the eigenstates of sub-units of the whole structure, producing an enhancement of the thermal coherent length. This represents a clear example of the interplay between structure and functionality. 
As a last remark, let us notice that even if  the other models (TD, PD)  have a super-transfer coupling between the ring eigenstates with the largest dipole strength,  the resulting SRS lies in the middle of the spectrum and its effect on the thermal coherence length
is less relevant (since the latter is sensitive to the density of states only in the lowest part of the energy spectrum).
 This argument strongly supports   the relationship between the presence of a SRS  close to the ground state  and the thermal coherence length   discussed above.

\section{Natural concentric structures}
\label{sec:conc}
 
\begin{figure}[t]
\centering
\begin{tabular}{cc}
{\bf \Large (A)} & {\bf \Large (B)} \\
\includegraphics[width=7.5cm]{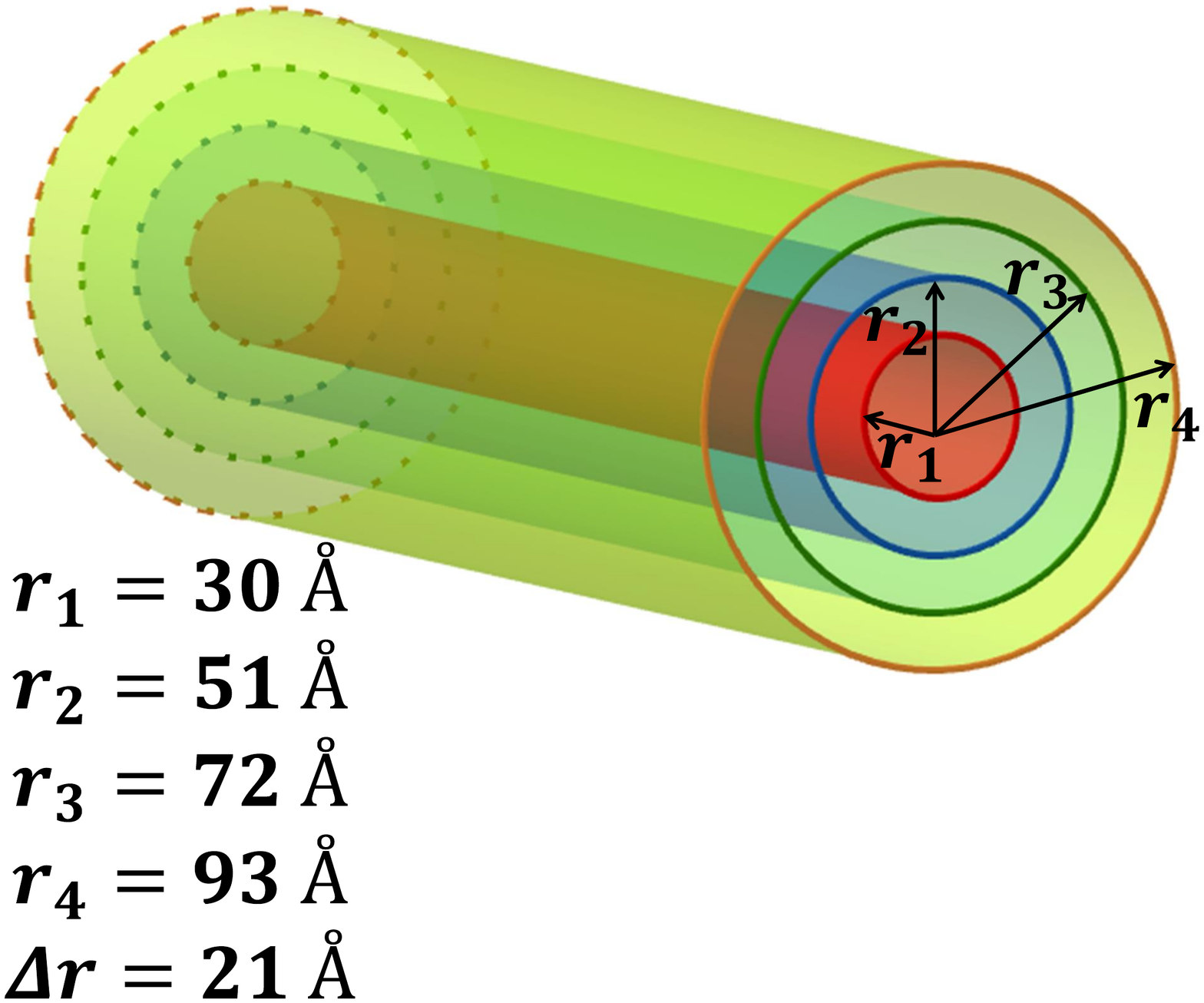} &
\includegraphics[width=7.5cm]{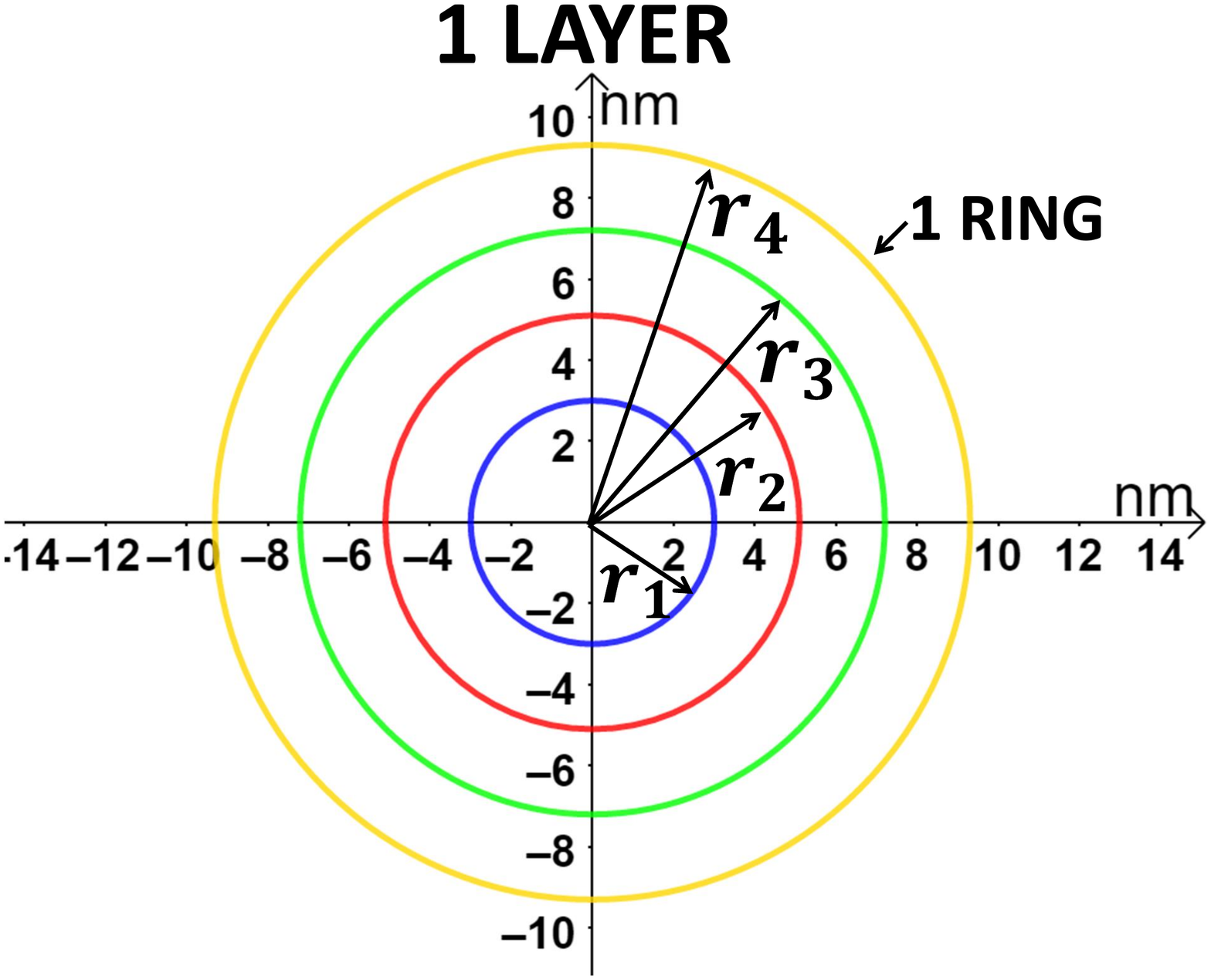} \\
\end{tabular}
\caption{(A) Structure of an aggregate of Bchl molecules on four concentric rolls. The radius of the innermost roll is $ 30 $ \AA, while the distance between consecutive layers is equal to $ 21 $ \AA.
(B) Single layer of the structure formed by four concentric rings. The whole aggregate has been obtained by overlying 100 layers~\cite{Huh}.}
\label{fig-4cyl}
\end{figure}

 Natural antenna complexes in Green Sulphur Bacteria are not made by a single cylindrical surface.
 In order to take this into account, 
 in this section we investigate a more complex configuration of dipoles on four concentric rolls as found in Green Sulfur bacteria \textit{Chlorobium Tepidum}.  Such structures have been extensively considered in literature (see for example \cite{Valleau,Koh,Saikin,Korppi}). Inspired from these studies we considered here a model of \textit{Chlorobium Tepidum} Triple mutant (bchQRU) formed by four concentric cylindrical surfaces, as shown in Figure~\ref{fig-4cyl}(A). Our aim is to investigate whether concentric cylindrical aggregates can support delocalized excitonic states at room temperature more efficiently than single cylindrical structures.

The distribution of the dipoles on each cylindrical surface  is the same as the MT model of the previous section.  
In Table~\ref{tab-01} we report all parameters for this model.
\begin{table}[!ht]
\centering
\begin{tabular}{|c|c|}
\hline
Number of surfaces & $ 4 $ \\
\hline
Radius of the innermost roll & $ 30 $\AA \\
\hline
Distance between concentric rolls & $21$ \AA \\
\hline
Radii of the cylinders & $ 30-51-72-93$ \AA \\
\hline
Number of dipoles on each ring & $ 30-51-72-93 $ \\
\hline
Density (number of dipoles over radius of the ring  \AA ) & $ 1$ constant value \\
\hline
\end{tabular}
\caption{Main parameters used to engineer the structure with four concentric rolls.}
\label{tab-01}
\end{table}

The coupling between the EMF and the dipoles of the aggregate
has been taken into account as in the Hamiltonian (\ref{eq:hreal}).
As in the previous sections let us first analyze the 
dipole strengths associated with the eigenstates of the Hamiltonian (\ref{eq:hreal}).
 
\begin{figure}[t]
\centering
\includegraphics[scale=0.6]{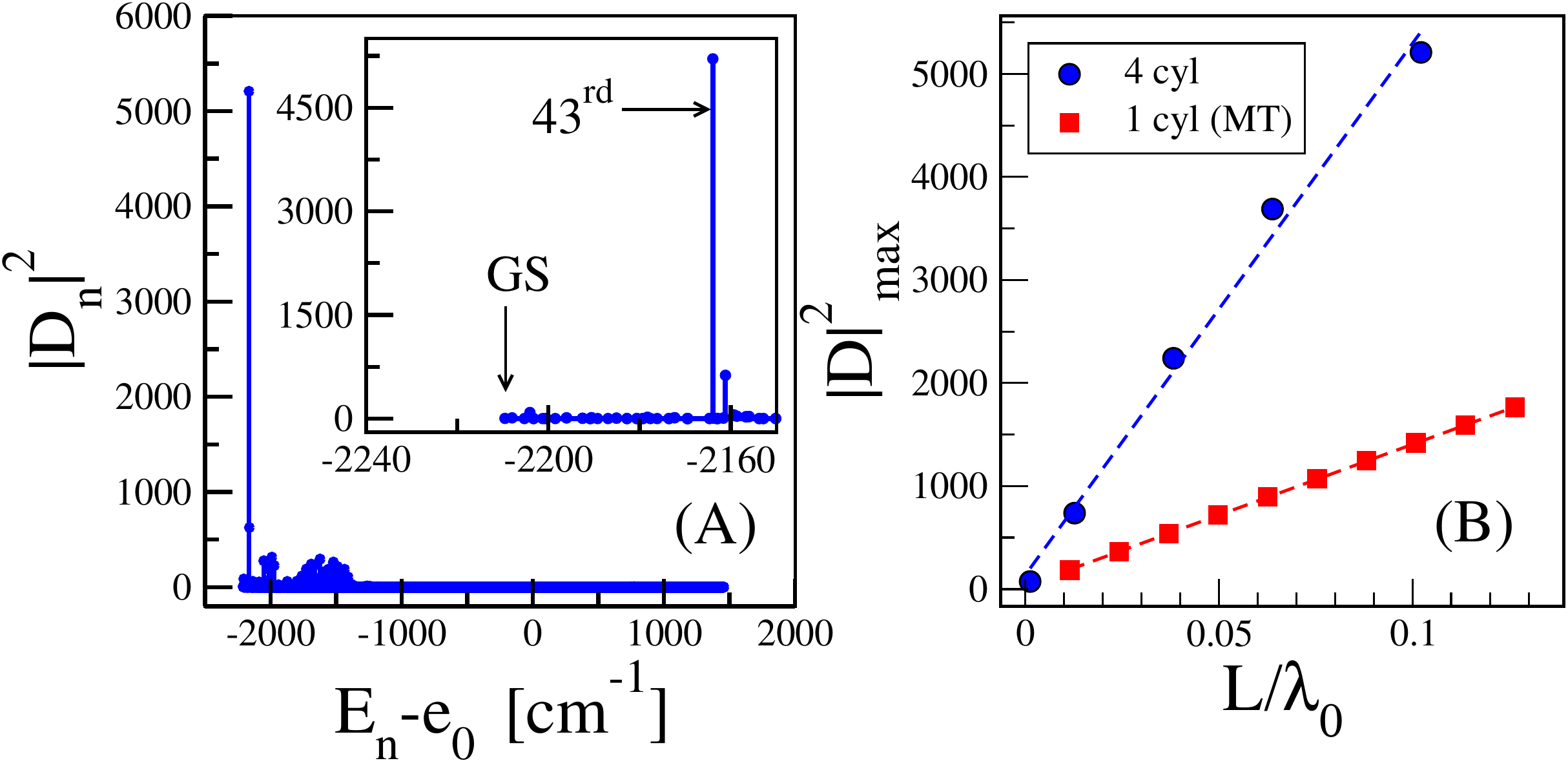}
\caption{(A) Dipole strength associated with each eigenstates of the system composed of 80 layers of 4 concentric rings for a total length of $L=65.57$ nm, as a function of the eigenvalues. Inset : the low energy part of the spectrum. Arrows indicate the Ground State (GS) and the state with maximal dipole strength (the $43^{rd}$ one).
(B) Maximal  dipole strength as a function of the rescaled  length of the aggregate $L/\lambda_0$ where $\lambda_0 \approx 650$ nm.  Dashed  line represent the linear fits. Maximal length considered in this panel is $L=65.57$ nm, corresponding to 80 layers of 4 concentric rings.  }
\label{dipole}
\end{figure}

Results are shown in Figure~\ref{dipole}(A) for a complex made of 80 layers of 4 concentric rings. As one can see the maximal dipole strength is concentrated in an energy region close to the ground state (the $43^{rd}$ eigenstate has the maximal dipole strength, see inset in  Figure~\ref{dipole}(A)).

Such dipole strength is associated with eigenstates   having  a high degree of delocalization along the cylinders. A further evidence is given in 
Figure~\ref{dipole}(B) where we show that the maximal dipole strength increases proportionally with the length $L$ of the cylinders. We also note that the maximal dipole  strength for concentric cylinders is between twice and 3 times larger than the maximal dipole strength of a single cylindrical surface with the same geometry, see  Figure~\ref{dipole}(B) where the same data of Figure~\ref{fig:secpar}(F) for the MT model have been reported for comparison. Note that the fact that concentric cylindrical surface can cooperate to create a larger SRS  is not trivial since the interaction between molecules in different cylinders is very weak, of the order of 16 cm$^{-1}$ which is one or two orders of magnitude smaller than the coupling between molecules inside each cylinder, see Table~\ref{tab-02}. 

Finally, we have studied the effect of  thermalization by putting  the system in a thermal bath at room temperature $ T=300 $ K. As before, we studied the thermal coherence length $ L_{\rho} $, see equation~(\ref{eq:lrho}).
 
\begin{figure}[t] 
\centering
\includegraphics[scale=0.6]{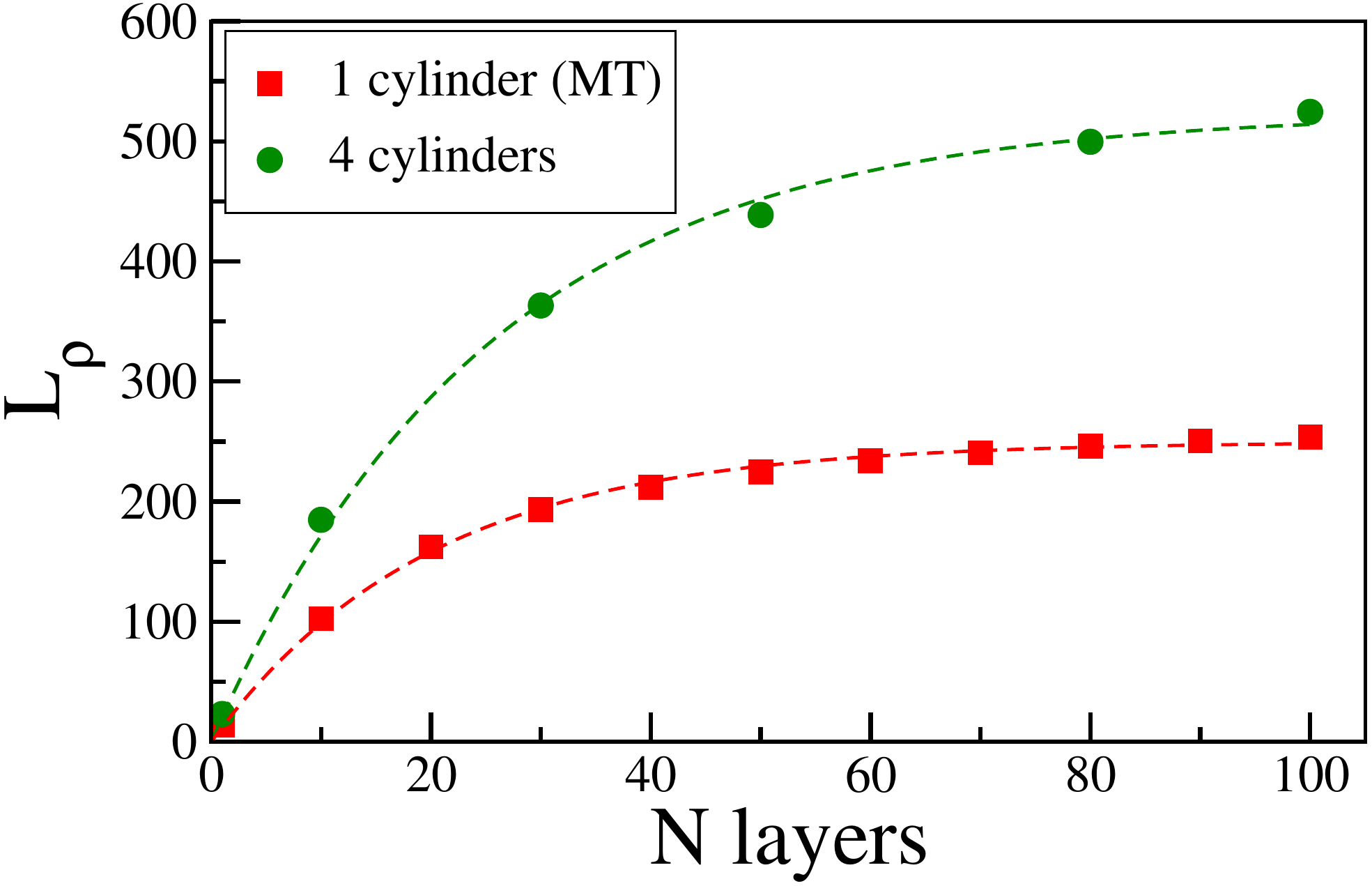}
\caption{ Thermal coherence length as a function of the number of layers in the cylinder for the system with four concentric cylinders (green circles) and the MT model with one cylinder only (red squares).
Dashed lines are the fit with the expressions~(\ref{eq:fit}) whose parameters are $L_{\infty}= 249.8$ and 
$N_c = 19.9$ for the dashed green curve
and $L_{\infty}= 523.9$ and 
$N_c = 25.2$ for the dashed red curve. }
\label{4cyl}
\end{figure}

 Results are shown in Figure~\ref{4cyl}(A) and compared with the same results obtained for the MT model. 
 A fitting with the function
 \begin{equation}
 \label{eq:fit}
 L_\rho = L_\infty \left( 1-e^{-N/N_c}\right),
 \end{equation}
 shown in figure as dashed lines, gives
 for the asymptotic coherence length (measured in number of layers) 
 $L_\infty = 532.9$ for the 4 cylinders and $L_\infty = 249.8$ for the MT model.
 Keeping in mind that the radius of the cylinder for the MT model is an average of the four radii of the structure composed of 4 concentric  cylinders, it is remarkable that the asymptotic coherence length is more than twice   larger than the single cylindrical structure.
This is highly non  trivial, since for the concentric cylinders we have many more states and the density of states is larger than that for the single cylinder having  the same length. For a discussion on this point see~\ref{app-c}. 

The results in this Section show that packing symmetrical structures in concentric cylinders  as it is found in natural photosynthetic complexes produces, at room temperature,  a  larger thermal coherence length than a single cylinder.

\section{Conclusions and Perspectives}
\label{conclu}
We have analyzed realistic structures of self-aggregated molecular nanotubes of chlorophyll molecules as  found in Antenna Complexes of Green Sulfur Bacteria. By taking into account position and dipole orientation of chlorophyll molecules which agree with experimental data we
have shown that natural structures are able to support macroscopic coherent states even at room temperature. Indeed in natural complexes we have found  delocalized thermal excitonic states with a  coherence length extending over hundreads of molecules. 
We show that such thermal coherence length is much  larger than that one could expect from the magnitude of the nearest-neighbour coupling and it cannot be explained even by the long-range nature of the interaction between the molecules. 
Instead, the  ability of natural structures to support a large coherence length can be traced back to their specific geometric features.

In order to explain how this is possible, 
we first considered cylindrical structures made of a sequence of rings, each containing a fixed number of molecules equally spaced on the ring itself. Since the disposition of the dipoles  is  highly symmetric, in each ring we have few superradiant eigenstates (to which we  associate a giant dipole) where most of the dipole strength of the system is concentrated, and many subradiant states with zero dipole strength. Moreover,  due to discrete rotational symmetry of the whole cylinder around its axis, each eigenstate of the ring sub-unit is coupled only with the correspondent eigenstate in the other rings. 
The coupling between the superradiant eigenstates in each ring gives rise to the super-transfer effect, i.e. a coupling which is enhanced by a factor proportional to the number of molecules in the ring. At the same time we have shown that the coupling between the subradiant states in each ring induces a sub-transfer effect, i.e. a suppressed coupling compared to  the single molecule coupling. 
Moreover, we have demonstrated that in natural complexes  the super-transfer coupling between the superradiant states in each ring    generates  the lower part of the energy spectrum of the whole cylinder. 

Since the spectral energy width of a system is proportional to the intensity of the coupling between its parts,  the enhanced super-transfer coupling is able to increase the spectral width close to the ground state. This creates a depressed density of states in the lower part of the spectrum, allowing for a larger thermal coherence length. Indeed  the latter
 increases as the number of states in an interval  $k_BT$ above  the ground state decreases.  We also gave evidence that similar mechanisms are responsible for the large thermal coherence length that we found in  other natural structures (WT model) where the disposition of the dipoles is less simple than the one described above. 
 
From our results we can predict that symmetry in cylindrical molecular nanotubes is essential to have robust structures, not only to thermal noise, as we considered in this paper, but also to other sources of  noise. The structural requirement is to create a super-transfer coupling between the superradiant eigenstates of cylindrical sub-units able to generate the lower part of the spectrum of the whole structure. 

Molecular nanotubes are fundamental structures in biological systems and they are among the most promising structures to be used in  quantum devices.  
The most important message which can be extracted from our analysis is the fact that specific geometric features, connected to symmetries, allow to control the cooperative effects in molecular aggregates. Indeed it is due to the presence of such cooperatively enhanced coupling  (super-transfer) inside the molecular aggregates that  macroscopic coherent states are allowed to survive at room temperature.  This is an emergent property of such structures which cannot be reduced either to the intensity of the coupling between the molecules, or to their interaction range.

The relevance of geometry in molecular aggregates and the emergent properties arising from it, are fundamental  to understand even more complicated structures. For instance,
structures made of few concentric cylinders as they are found in Green Sulphur bacteria. Our preliminary study of such structures has  shown that these aggregates have an enhanced thermal coherence length compared to the single cylindric surfaces. 
We would like to mention that recently by some of the  authors of this paper, excitonic states have been  analysed also in Microtubules~\cite{Phil},  which are molecular nanotubes thought to be involved in many cellular functions.
The analysis have confirmed the role of symmetry and geometry in such structures too. 

In perspective it would be interesting to investigate the relevance of macroscopic coherent states for light-harvesting and photo-excitation transport. Moreover it would be important to 
understand the general structural requirements necessary to induce macroscopic coherent states in generic molecular networks.

\ack  
GLC acknowledges
the support of PRODEP (511-6/17-8017). Useful discussions with J. Knoester, N. Keren and G. G. Giusteri are also acknowledged.  

\newpage

\appendix

\section{Geometry of the models}
\label{app-a}
\begin{figure}[t]
\centering
\includegraphics[scale=0.35]{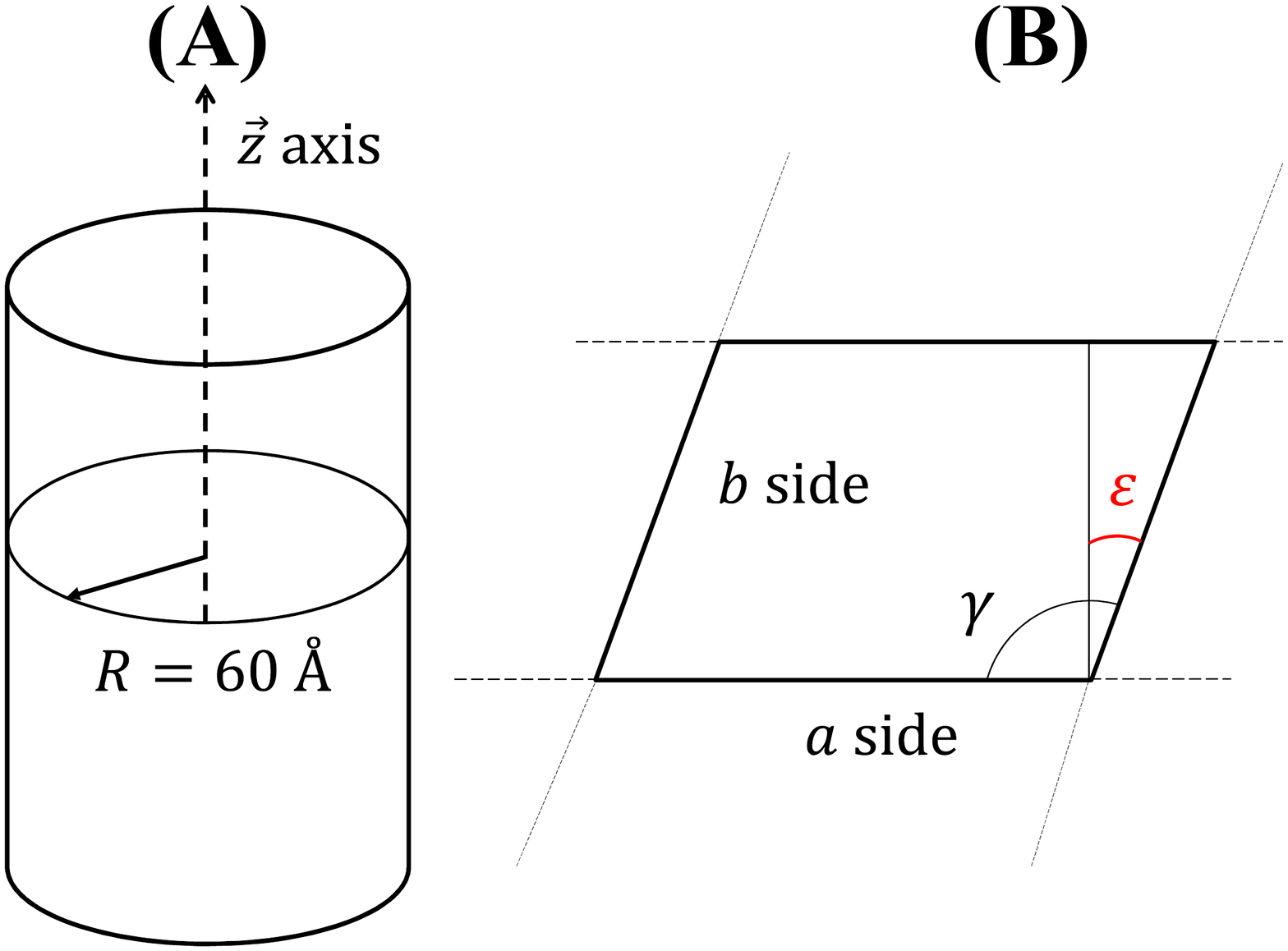}
\caption{(A) Schematic  cylindrical structure of each model. The cylindrical axis corresponds to the $\vec{z}$ axis and the radius is    $R=60 \mbox{ \AA}$. (B) Fundamental unit cell of the analyzed aggregates. One may obtain each of the models varying the three parameters $a$, $b$ and $\gamma$.   In our models we consider only the cases in which $a$ and $\vec{z}$ are orthogonal or parallel.}  \label{fig:cyl}
\end{figure} 
We analyzed  five different
cylindrical models with  fixed radius ($R=60 \mbox{ \AA}$) and  total number of chromophores $N$, as  shown in Figure~\ref{fig:cyl}(A). These models differ for  the geometrical arrangement of the chromophores (dipoles) along the cylindrical surface. 

In order to describe how the dipoles are placed on the cylindrical surface let us wrap it up  on a rectangular plane. In this plane the dipoles are disposed on the vertices of a lattice.
The unit cell of the  lattice in all structures is created  starting from two lattice parameters $a$ and $b$ arranged in such a way  that the angle between them is $\gamma$, see Figure~\ref{fig:cyl}(B).
  Depending on the particular arrangement of the unit cell in the  lattice the dipoles can be arranged into vertical chains or placed onto equal horizontal and coaxial rings. 
We assume that,  in each of these structures,  the shortest distance between two chromophores located on the same ring is  ${r}_1=6.28 \mbox{ \AA}$.  
In our  scheme  all chromophores can be treated as dipoles with a constant squared dipole moment $|\mu|^2=30 \mbox{ D}^2$ \cite{Holz}. This   corresponds to a  dipole length  $L_d=1.14 \mbox{ \AA}$ \cite{Calculus}. The ratio $L_d/{r}_1 \simeq 0.18$  is relatively small so that the dipole approximation can be successfully applied for the nearest-neighbor coupling. \\  

In the following subsections we will analyze in details the  geometrical structures associated with each model.
\subsection{MT model}
\begin{figure}[t] 
\centering
\includegraphics[scale=0.35]{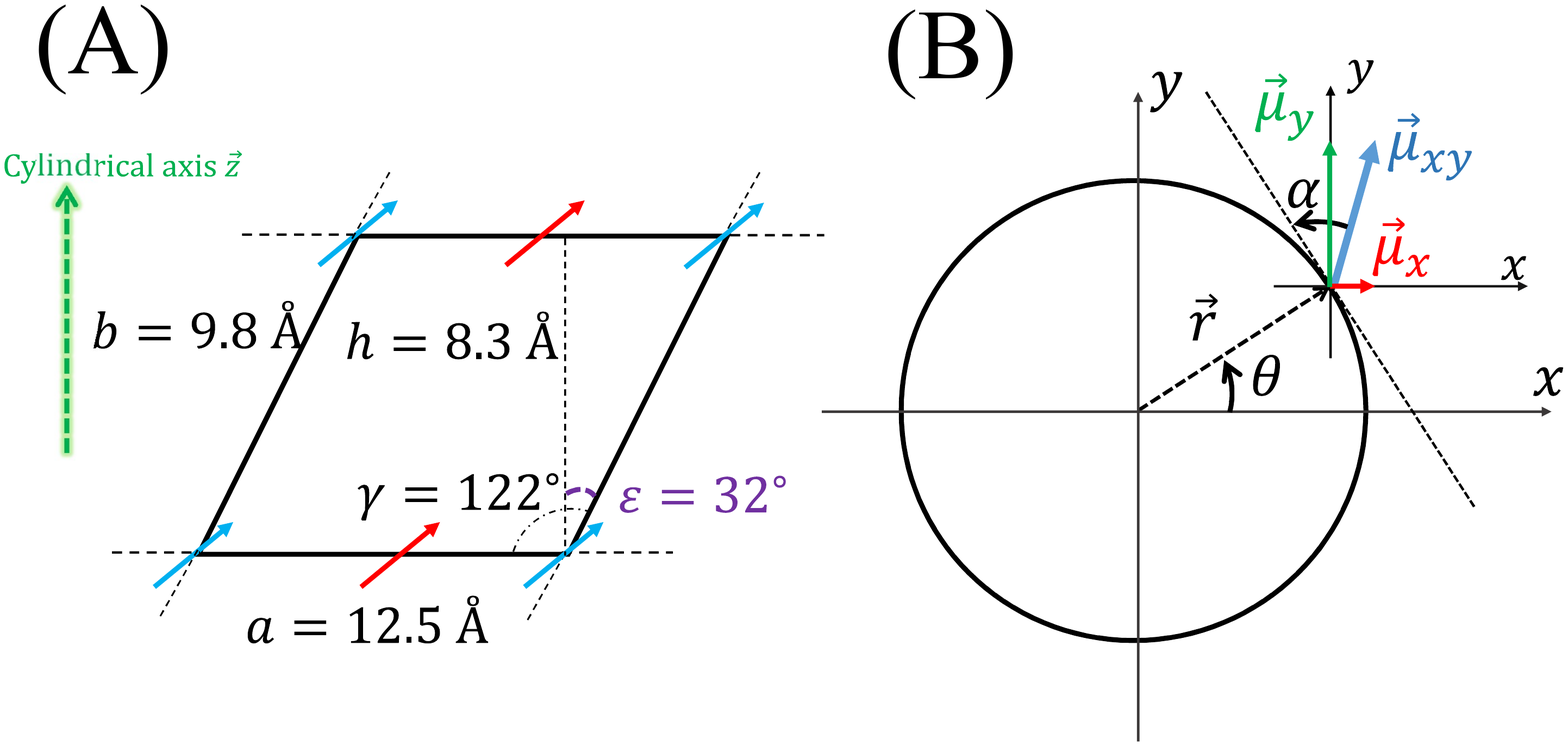}
\caption{ 
(A) Unit cell for the MT model. Here $h=8.3 \mbox{ \AA}$ is the vertical distance between two consecutive rings. The cylindrical axis $\vec{z}$ is represented by the green arrow on the left and is perpendicular to the $a$ side. Note that the alternation between the colours of two consecutive dipoles along the $a$ and the $b$ sides is due to the alternation $\alpha=\pm 4^\circ$.(B) View from above of the single ring of the MT type. Projection $\vec{\mu}_{xy}$ is the projection in the $xy$ plane of the dipole $\vec{\mu}$.
}
 \label{fig:mtsect}
\end{figure}
%
The MT model proposed here coincides  with the Chlorobium Tepidum bchQRU triple mutant investigated in other studies (\cite{Ganapathy,Koh,Chew}).
\\ In the MT cylindrical structure, the total number $N$ of chromophores is organized into $N_1$ equal, horizontal and coaxial rings, see Figure~\ref{fig:cyl-all}(A) in the main text.  Each ring contains $N_2=60$ chromophores and two consecutive rings are separated by a vertical distance $h=8.3 \mbox{ \AA}$. In the unit cell, shown in Figure~\ref{fig:mtsect}(A),  $h$ is parallel while $a$ is perpendicular
to the cylindrical axis $\vec{z}$.
Any  chromophore along the surface will be labelled  as the $n_2^{th}$ dipole on the $n_1^{th}$ ring (where $n_1=1,...,N_1$ and $n_2=1,...,N_2$). Since we keep the radius $R$ fixed, the density of chromophores along each ring  is also fixed: $\rho_s= (2\pi \mbox{\AA})^{-1}$.
The position of each  dipole onto the cylindrical surface is characterized by two cylindrical coordinates. Nevertheless it 
is useful to introduce three angles (the latter being dependent on the first and second):
\begin{itemize}
\item $\varphi=360^\circ/N_2=6^\circ$ is the azimuthal angle between two adjacent dipoles in the same ring,
\item $\xi=h \tan \varepsilon/R   \simeq 4.956^\circ$ is the shift angle between two successive dipoles located onto neighbour rings,
\item $\theta=n_1\xi+n_2\varphi$ is the angle between the position  $\vec{r}$ of the dipole and the $x$ axis. 
\end{itemize}
In this way we have :
\begin{eqnarray} \label{eq:psmt}
r_x & = & R \cos \theta \nonumber \\
r_y & = & R \sin \theta \\
r_z & = & hn_1 \nonumber
\end{eqnarray} 
The components of the dipole moment $\vec{\mu}$ can be expressed through two angles:
\begin{itemize}
\item $\alpha=4^\circ$,  between  the projection of the dipole moment onto the plane of the ring  and the plane tangent to the cylindrical surface, see Figure~\ref{fig:mtsect}(B).
\item $\beta=55^\circ$, is the angle created by the single dipole moment with the cylindrical axis.
\end{itemize}
Assuming the $n_2^{th}$ dipole with an angle $\alpha=+4^\circ$, the $(n_2+1)^{th}$ dipole will have $\alpha=-4^\circ$, the $(n_2+2)^{th}$ dipole   $\alpha=+4^\circ$ and so on. This alternation is valid along the $a$  direction and makes consecutive dipoles to point inward ($\alpha=+4^\circ$) and outward ($\alpha=-4^\circ$) respectively. Generally, we have that the generic dipole moment $\vec{\mu}$ has the following normalized components expressed in terms of spherical coordinates:
\begin{eqnarray} \label{eq:dpmt}
\mu_x & = & -\sin \beta \sin (\theta + (-1)^{n_2} \cdot \alpha) \nonumber \\
\mu_y & = & \sin \beta \cos (\theta + (-1)^{n_2} \cdot \alpha) \\
\mu_z & = & \cos \beta \nonumber
\end{eqnarray}
%

\begin{figure}[t]
\centering
\includegraphics[scale=0.3]{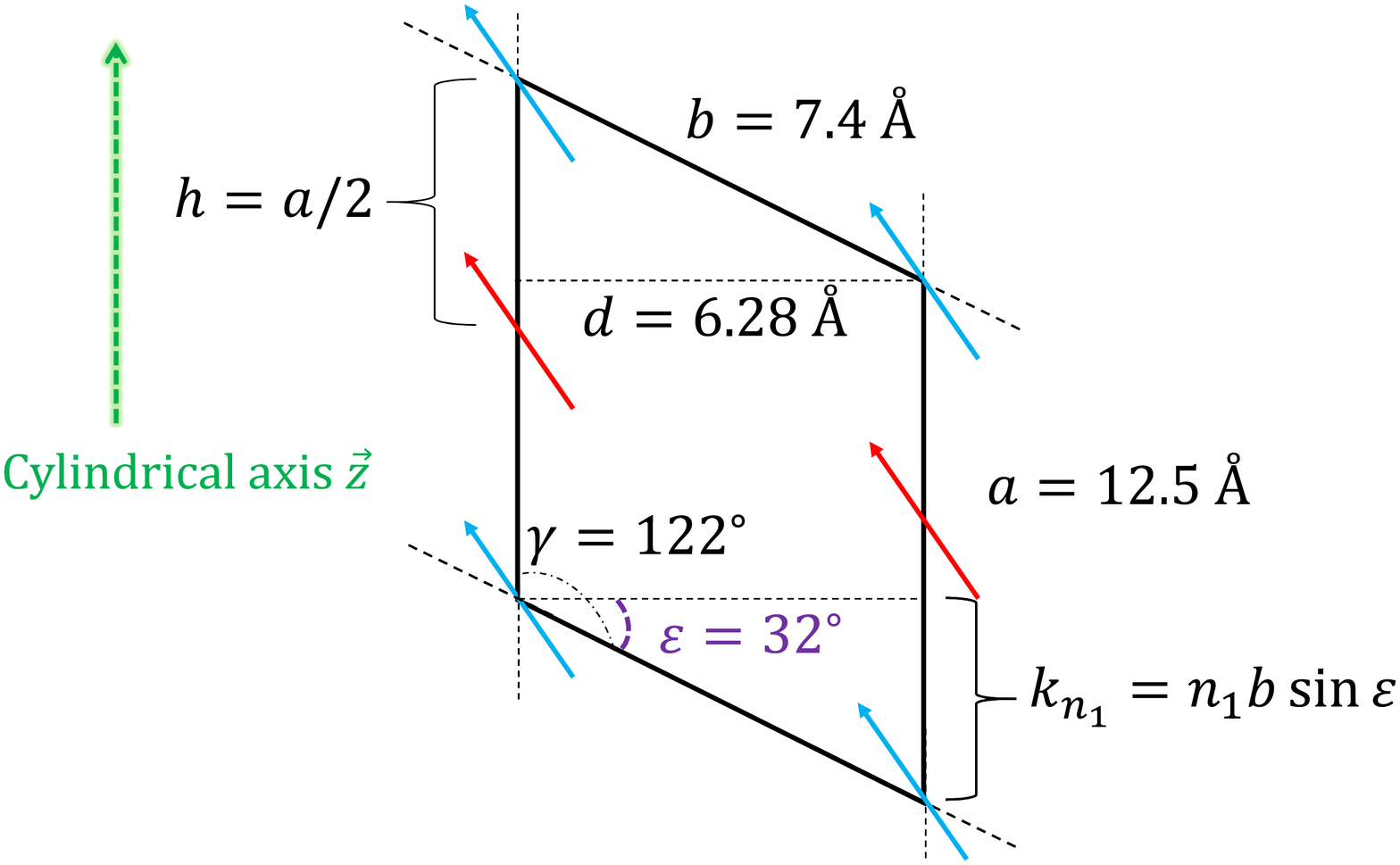}
\caption{ Unit cell for the WT model. Here $k_{n_1}$ refers to the vertical shift of two dipoles belonging to two nearest-neighbour chains, while $d$ is the distance between two chains. Two consecutive dipoles on the same chain are separated just by a quantity equal to $a/2$. One can notice that this unit cell is the rotation by 90$^\circ$ around $\vec{z}$ of the previous unit cell for the MT model (Figure \ref{fig:mtsect}), but with different parameters. The alternation of two colours of two consecutive dipoles along the $a$ and the $b$ direction represents the typical alternation $\alpha=\pm 4^\circ$.
} 
\label{fig:wtlatt1}
\end{figure}
 
\subsection{WT model} 
The WT model~\cite{Ganapathy,Koh,Chew} 
 shows a deep structural difference compared to the other structures.
Indeed it can be thought as  organized into  $N_1$ vertical chains and  each of them with $N_2$ molecules. So one can talk about the $n_2^{th}$ chromophore on the $n_1^{th}$ chain with $n_1=1,...,N_1$ and $n_2=1,...,N_2$.  Dipoles moments on adjacent chains do not have the same height but they are shifted by a quantity $k_{n_1}=n_1 b \sin \varepsilon$ to originate a helical structure, as shown in Figure~\ref{fig:cyl-all}(B) of the main text. The lattice has the following parameters: $a=12.5 \mbox{ \AA}$, $b=7.4 \mbox{ \AA}$, $\gamma=122^\circ$ and $\varepsilon=32^\circ$. The unit cell of the WT type is similar to that of the MT, but it is rotated by an angle $90^\circ$ around the $\vec{z}$ direction, so the vertical distance between two dipoles on the same chain measures $h=a/2=6.25 \mbox{ \AA}$. Moreover 
$\varphi=360^\circ/N_1=6^\circ$ will be intended as the azimuthal angle between adjacent chains, and $\theta=n_2\varphi$ as the angle between the position vector $\vec{r}$ and the $x$ axis. 
The position of the generic dipole on the surface can be expressed in cylindrical coordinates as follows: 
\begin{eqnarray}
r_x & = & R \cos \theta \nonumber \\ 
r_y & = & R \sin \theta \\ 
r_z & = & hn_2+k_{n_1} \nonumber 
\end{eqnarray} 
The components of each dipole moment  are given by  equation~(\ref{eq:dpmt}) with  $\alpha=4^\circ$, $\beta=35^\circ$.
Also in the WT model there is  the alternation $\alpha=+4^\circ$ and $\alpha=-4^\circ$ between consecutive dipoles.

\begin{figure} [t]
\centering
\includegraphics[scale=0.3]{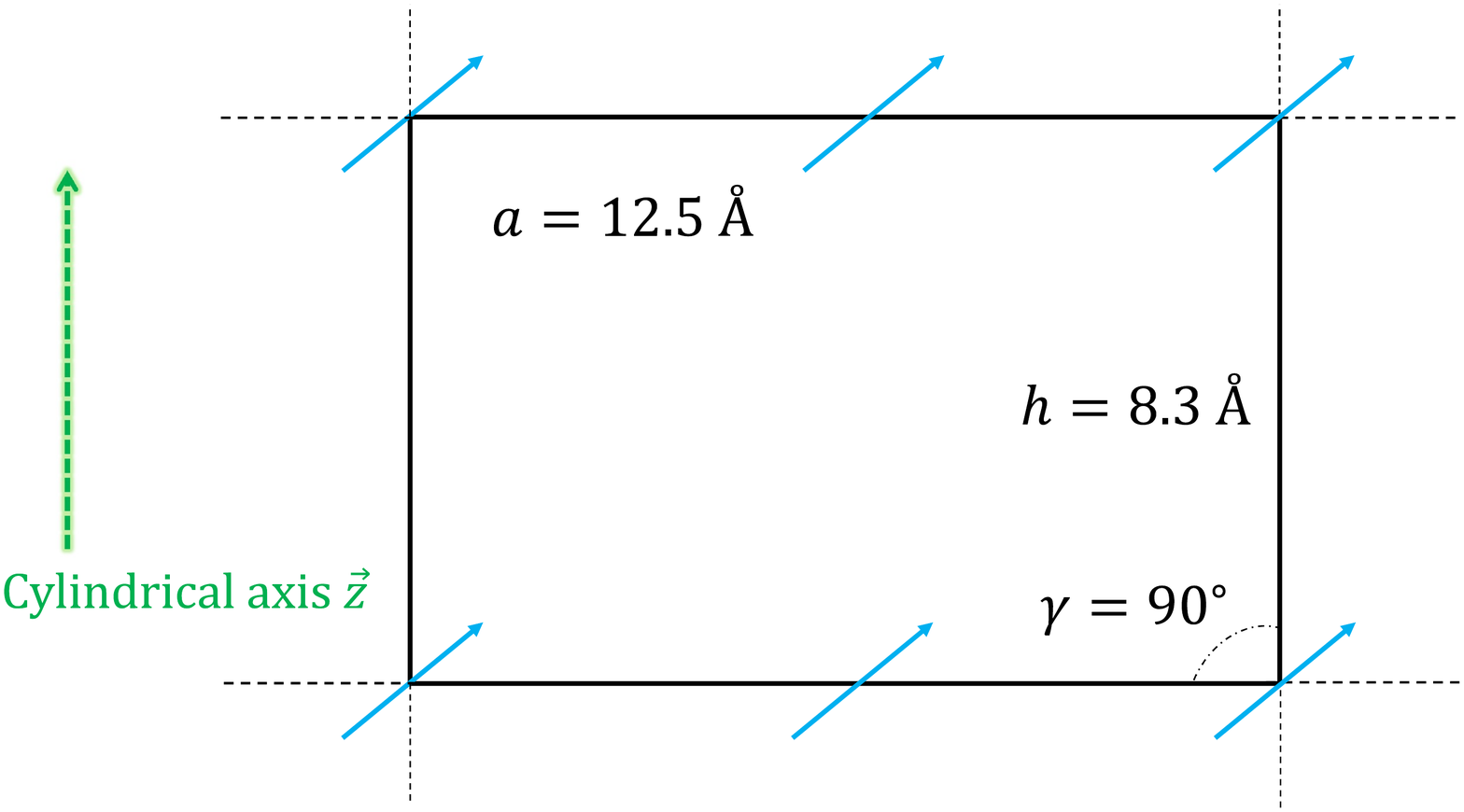}
\caption{ Unit cell for the PD, TD and RD models. Here the parameter $b$ coincides with the vertical distance $h$ between two consecutive rings. The cylindrical structure is obtained wrapping this rectangular unit cell around the direction of $\vec{z}$, which is perpendicular to the $a$ side.

}
\label{fig:pdm}
\end{figure} 
\subsection{PD, TD and RD models}
These models, shown in  Figure~\ref{fig:cyl-all}(C, D, E) of the main text  do not exist in nature and  they have been introduced only for comparison with the natural systems.
They exhibit a different lattice compared to the natural complexes, since the unit cell   is a rectangle.
 As shown in Figure~\ref{fig:pdm} we have  $\gamma=90^\circ$,   $a=12.5 \mbox{\AA}$ and   $b=h=8.3 \mbox{ \AA}$. 
 The cylindrical axis $\vec{z}$ is perpendicular to the $a$ side 
and one could build each of the three structures wrapping up the  lattice around it. The three cylinders have again the same
 radius $R=60 \mbox{ \AA}$ and the same  number of molecules $N$. Also, they are  arranged into $N_1$ equal, horizontal and coaxial rings such that each of them carries $N_2$ dipoles. Once again a particular dipole moment will be indicated as the $n_2^{th}$ chromophore on the $n_1^{th}$ ring ($n_1=1,...,N_1$ and $n_2=1,..., N_2 $).  The three models differ for the values of $\alpha, \beta$ in the following way: \begin{itemize}
\item in the PD structure, $\beta=0^\circ$ and $\alpha=0^\circ$,
\item in the TD structure, $\beta=90^\circ$ and $\alpha=0^\circ$, 
\item in the RD structure, the angles are uniformly distributed such that $\alpha \in [0,2\pi]$ and $\beta \in [0,\pi]$.
\end{itemize}

\section{Comparison between dipole strengths and radiative decay widths}
\label{app-b}

 \begin{figure}[t]
\begin{center}
\includegraphics[scale=0.6]{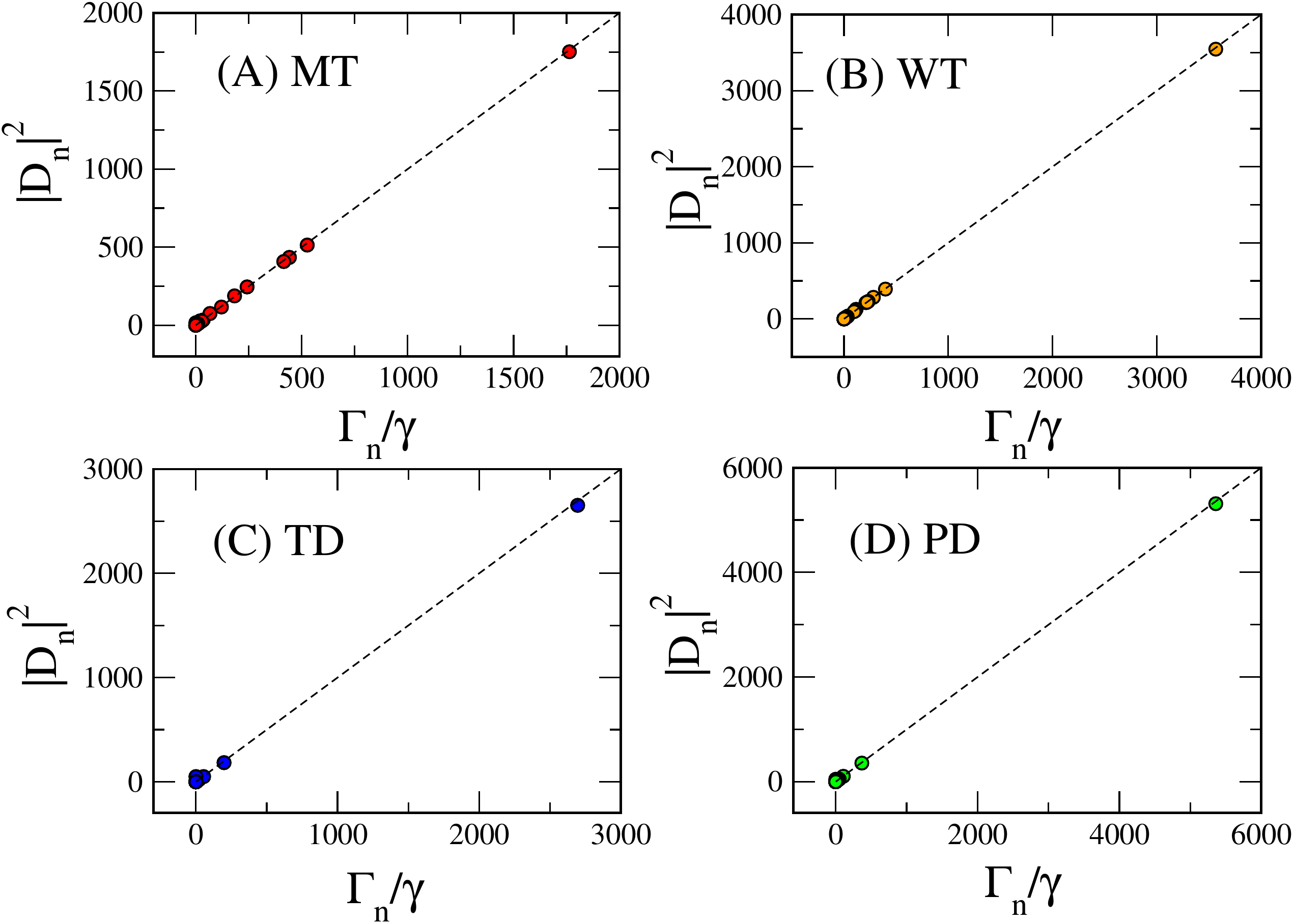}
\caption{(A) - (D) Squared dipole strength $|D_n|^2$ obtained using the real Hamiltonian $H_r$~(\ref{eq:hreal}) as a function of the radiative decay $\Gamma_n/\gamma$ obtained diagonalizing the Non-Hermitian Hamiltonian $H$~(\ref{eq:ham}), for a total number of dipoles $N=6000$. The trend is manifestly linear in each model as it can be noticed from the superposition with the line $|D_n|^2=\Gamma_n/\gamma$. This matter of fact confirms that the imaginary part of the   Hamiltonian given in equation~(\ref{eq:ham}) of the main text is perturbative indeed. Using the real Hamiltonian equation~(\ref{eq:hreal}) of the main text  would not result in any significant difference.} 
\label{fig:linears}
\end{center}
\end{figure}
 
In Figure~\ref{fig:linears} the comparison between 
the dipole strengths $|D_n|^2$ (obtained using the Hamiltonian $H_r$~(\ref{eq:hreal})) and the rescaled radiative widths $\Gamma_n/\gamma$ (obtained diagonalizing the Non-Hermitian Hamiltonian $H$~(\ref{eq:ham})) is shown for all the eigenstates of the MT, WT, TD and PD models for $N=6000$ dipoles. As one can see the two quantities can be considered to be the same (compare symbols with the dashed lines, which represent the behaviors $|D_n|^2=\Gamma_n/\gamma$).
\section{Comparison between radiative and dipole approximations}
\label{app-bc}
\begin{figure}[t]
\centering
\includegraphics[scale=0.5]{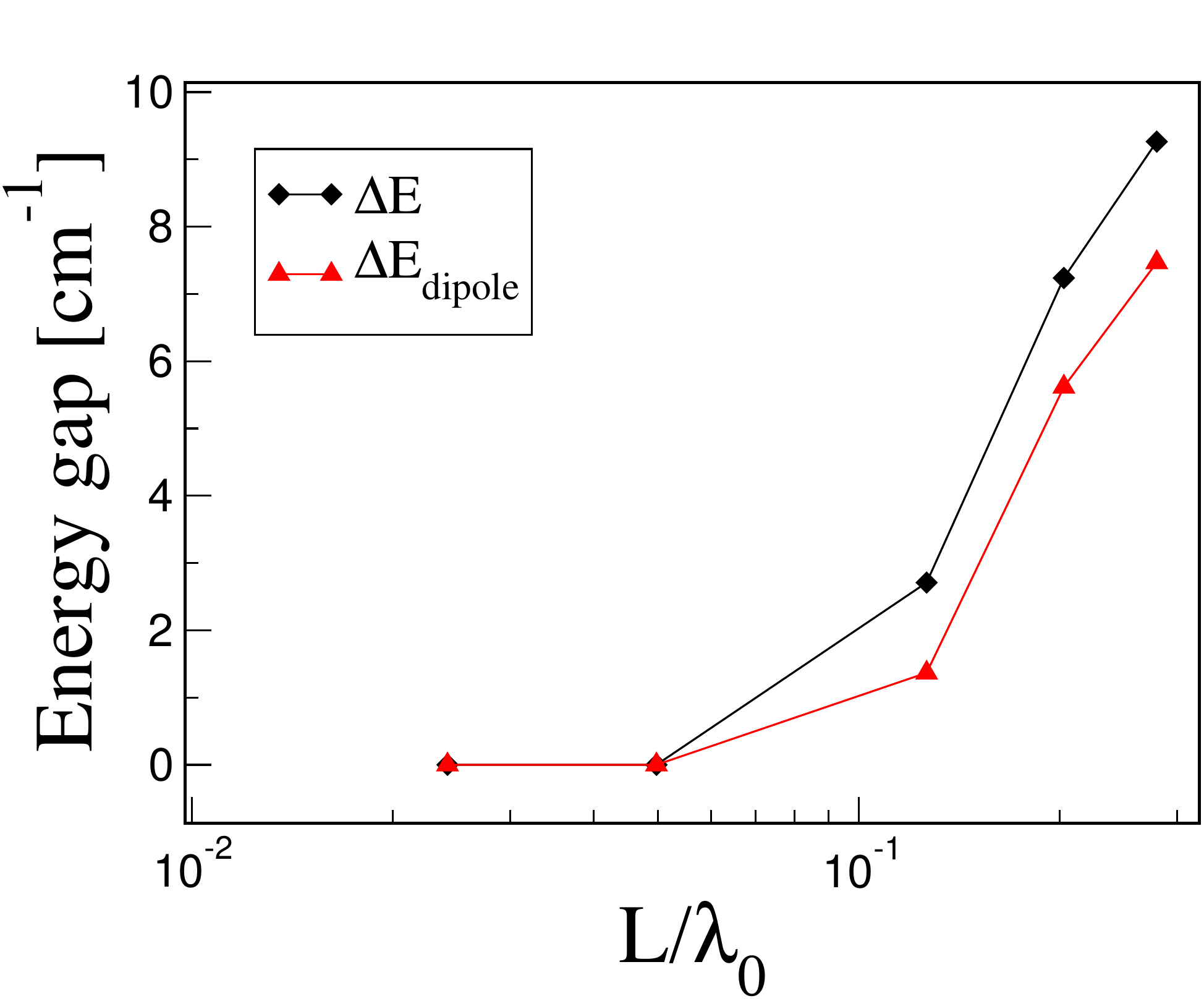}
\caption{Energy gap between the ground state and the first excited state, computed for the radiative Hamiltonian (black squares), see equation~(\ref{eq:hreal})  in the main text,  and the dipole Hamiltonian (red triangles), see equation~(\ref{real}) in the main text.
}
\label{gapA}
\end{figure}

In order to understand the validity of the dipole approximation in the range of sizes of the natural systems considered, 
we have compared the dipole Hamiltonian, see equation~(\ref{real}) in the main text, with the radiative Hamiltonian, see equation~(\ref{eq:hreal})  in the main text, which we used in our paper. For instance comparing the dipole strength and the energy of the superradiant state we have found that the dipole approximation is good for both quantities, with a relative error which increases with the system size, but it remains small up to the value of $L/\lambda_0 \approx 0.3$ where the relative error of the dipole strength is $0.1 \%$ and the relative error of the  energies is $0.02 \%$. Nevertheless in other quantities, such as the energy gap between the ground state and the first excited state, the error  can be as large as  $20 \%$, see Figure~\ref{gapA}. Thus, we can say that while the dipole approximation seems to be well justified for the typical sizes of natural nanotubes, nevertheless non-negligible deviations can be found in some relevant quantities. For this reason here we used the radiative Hamiltonian which is more accurate. Moreover, one should not forget that the errors increase with the system size.

\section{Participation ratio of the eigenstates}
\label{app-pr}
\begin{figure}[t]
\centering
\includegraphics[scale=0.6]{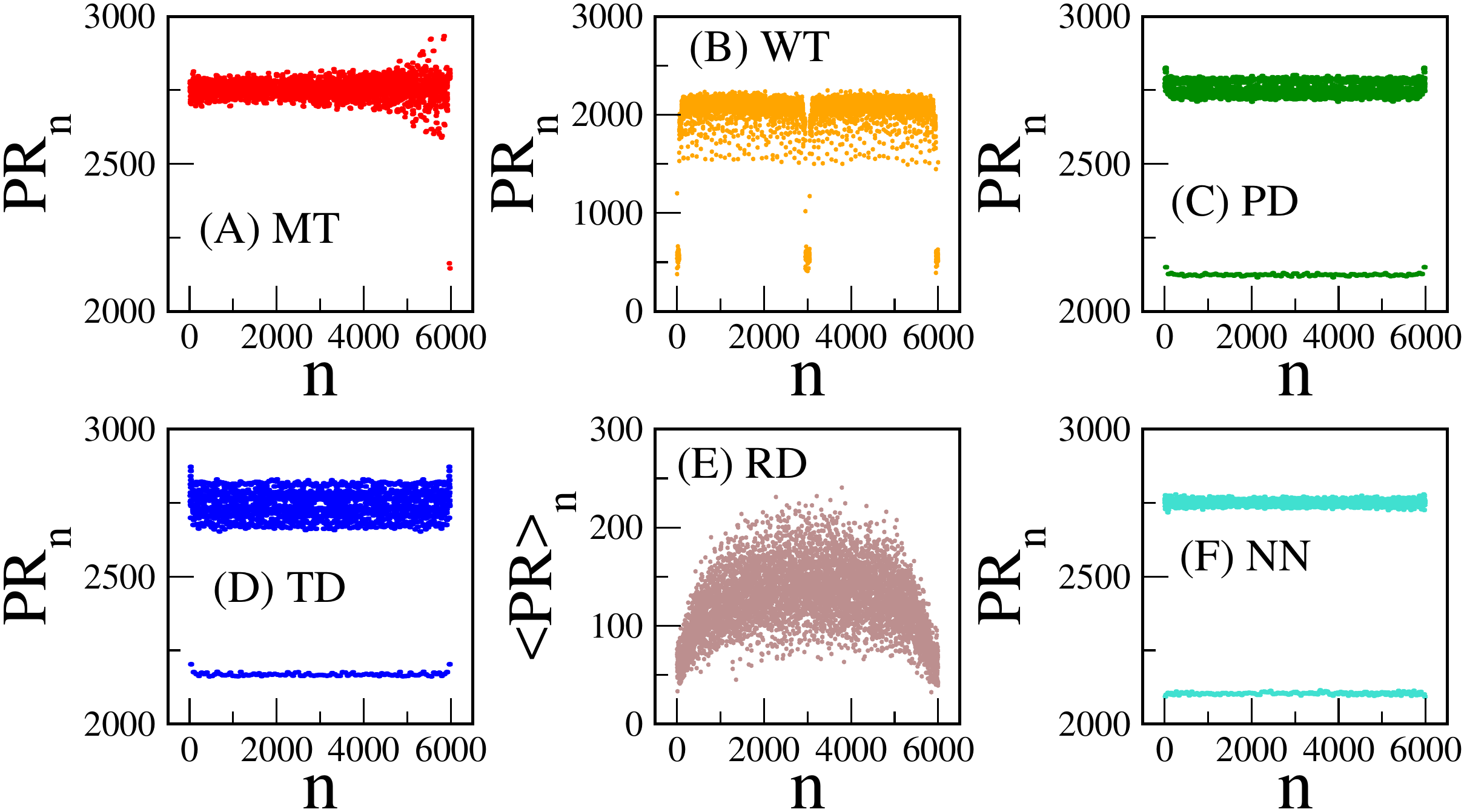}
\caption{(A) - (F) PR of the eigenstates as a function of the eigenstate index. Most structures, both natural and artificial, show a PR of the same order of magnitude of the total number of dipoles $N$. An exception to this trend is represented by the RD model (panel (E)), in which the PR is smaller of about one order of magnitude. Note that in this case we speak of $\langle PR \rangle$, since the PR has been calculated for 10 disorder realizations. In all cases we considered
  $N=6000$ dipoles.}
\label{fig-pr}
\end{figure}

As a measure  of delocalization of the eigenstates of the different nanotubular structures, we analyze the participation ratio (PR) of the eigenstates. Let us take into account the expression equation~(\ref{eq:expan}) of the $n^{th}$ energy eigenstate on the site basis: the coefficient $C_{ni}$ indicates  its component  on the $i^{th}$ site. The PR of the $n^{th}$ eigenstate is defined as follows:
\begin{equation} \label{eq:IPR}
PR_{n}=\frac{1}{\sum_{i=1}^{N} |C_{ni}|^4}. 
\end{equation}
Generally speaking, $PR_{n}\sim o(N)$ stands for a suitable degree of delocalization of the $n^{th}$ eigenstate, while we have $PR=1$ for a state fully localized  on a single site. Figure~\ref{fig-pr} shows how the PR of each eigenstate depends on the eigenstate index  in the six cylindrical models examined so far.  All models but the RD (E) exhibit a participation ratio of the same order, such that $PR_{n}\sim o(N)$. One may observe indeed a difference of about one order of magnitude between the RD aggregate and the other structures. The presence of low degree of delocalization in the RD model is expected since the  random coupling matrix elements between   molecules  can   induce Anderson localization~\cite{Anderson}. 
 
Note that the expression of   $L_{\rho}$  in equation~(\ref{eq:lrho}) is not equivalent to the PR even for the case of a density matrix describing a pure state.
Nevertheless  both  $L_{\rho}$  and the $PR$ are a measure of delocalization. 
 
\section{Super-Transfer in the WT model}
\label{app-wt}

\begin{figure}[ht!]
\begin{center}
\includegraphics[scale=0.5]{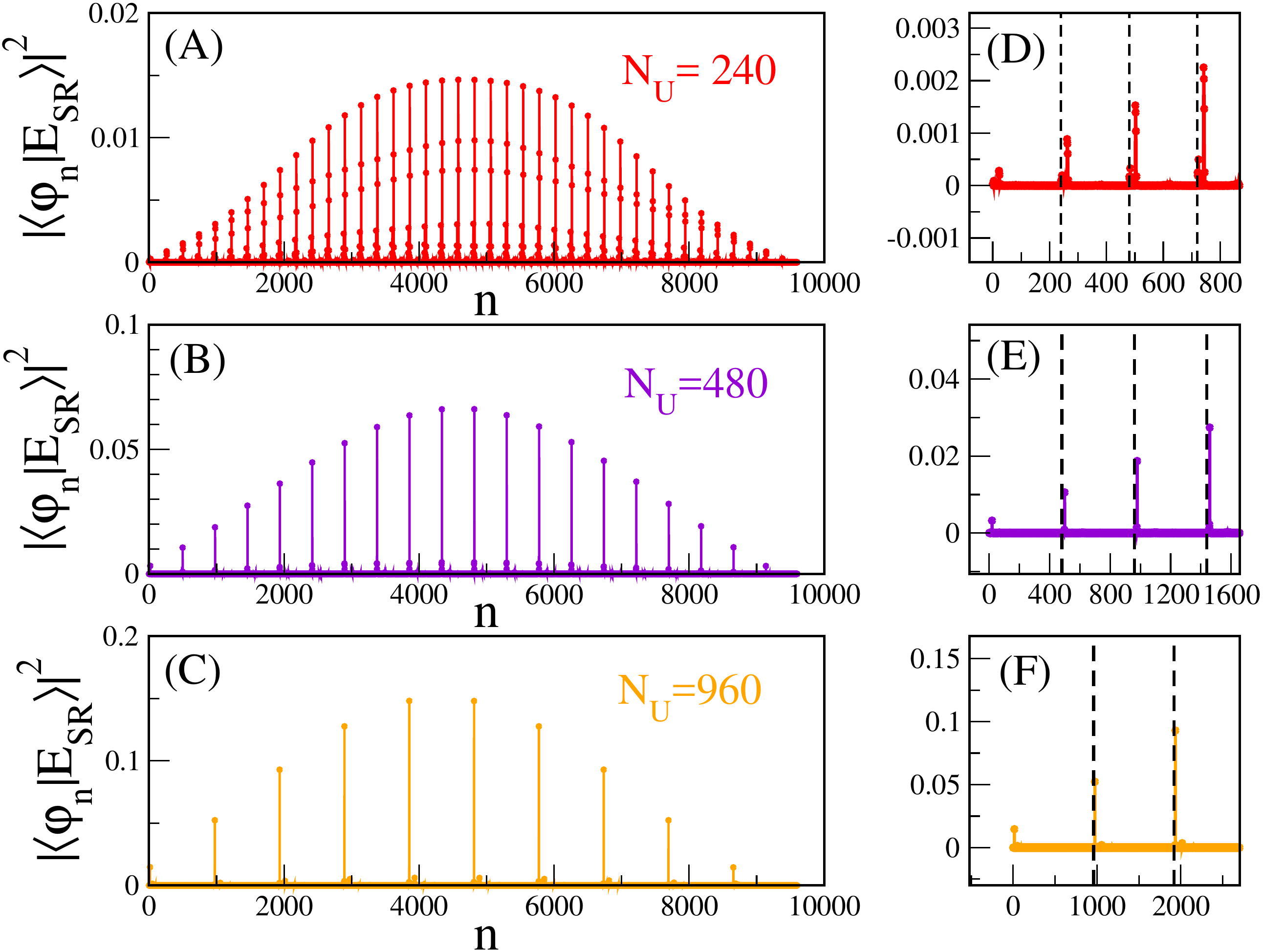}
\caption{ Projections of the most SRS  of the whole cylinder $\ket{E_{SR}}$ for the WT model composed of 9600 dipoles over the eigenstates $\ket{\varphi_n}$ of smaller cylinders composing the whole one. The smaller cylinder has been obtained by dividing the whole cylinder in smaller sub-units along its main axis length. The length of the sub-units has been varied as follow : $N_{U}=240$ (A),   $N_{U}=480$ (B) and  $N_{U}=960$ (C).  In the case considered in this figure the SRS corresponds to the second excited state $|E_{\rm 3}\rangle$. Panels (D,E,F) are   enlargements of (A,B,C) respectively. The vertical dashed lines indicates where each sub-unit ends.  
} \label{wtproj}
\end{center}
\end{figure}
The WT model is more complicated than the other models since the dipoles are not arranged into rings, but rather into helical structures. Nevertheless a very highly symmetrical disposition of the dipoles is also present in this case and one can think that the super-transfer coupling between the eigenstates of sub-units of the  whole cylinder might influence the lowest part of the spectrum even for this model. 
To show that we have split the whole cylinder along the axis direction (the $z$ direction) into smaller cylindrical structures. Each smaller cylinder contains a variable number of dipoles $N_{U}$.
The projection of the SRS of a cylinder of $9600$ dipoles on the eigenstates of these sub-units is shown in Figure~\ref{wtproj} for different values $N_{U}=240,480,960$. For a WT cylinder of $9600$ dipoles the SRS lies in the lower part of the spectrum and it corresponds to the second excited state with a dipole strength directed along the $z$-axis. As one can see from  Figure~\ref{wtproj} the SRS of the whole cylinder has components mainly on one eigenstate for each sub-unit. We checked that such eigenstate corresponds to a superradiant state SRS of each sub-unit with a dipole strength directed along the z-axis. Since the SRSs of each sub-unit have a giant dipole strength they  couple by super-transfer. This shows that also for the WT model the super-transfer coupling inside the cylindrical structure might be responsible for the low density of states  close to the ground state energy, see Figure~\ref{fig:lrho}(B). Nevertheless further analysis is needed to confirm this conjecture for the WT model.

\section{Concentric cylinders}
\label{app-c}

\begin{figure}[ht!]
\centering
\includegraphics[scale=0.6]{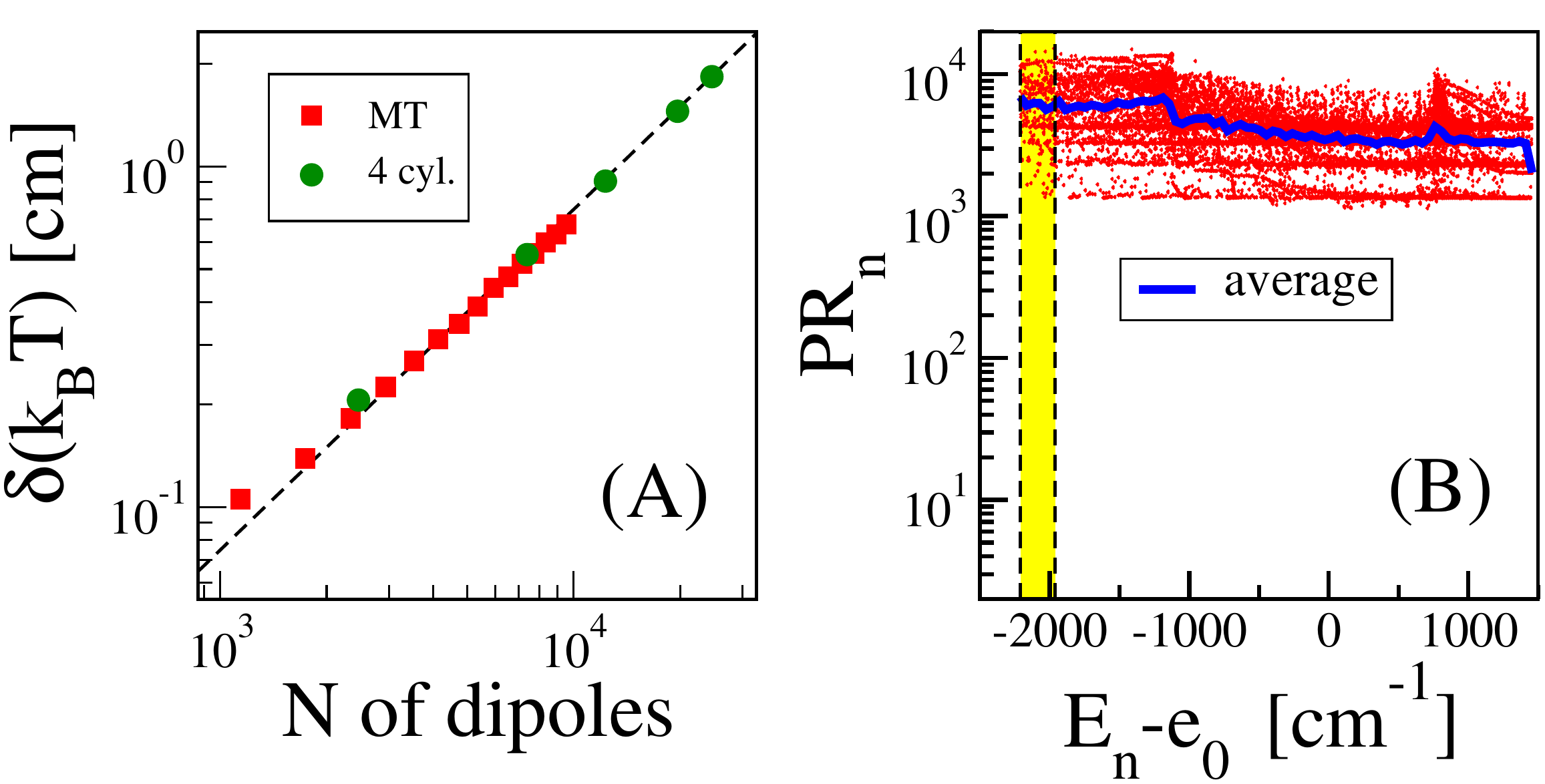}
\caption{(A) Density of states (number of states per unit  of thermal energy $k_B T$ at room temperature $T=300 K$, as a function of the total number of dipoles for the (MT) model (red squares) and for the four concentric cylinders (green circles).
Dashed line stands for $\delta (k_B T)  \propto N$.
(B) Participation ratio as a function of the shifted eigenenergy (red points). Blue curve represents the average.
Yellow region indicates the width of the thermal region $k_B T$.
}
\label{pr4}
\end{figure}

Let us emphasize that
the fact that the coherence length for the four concentric cylinders is larger than the single cylinder is highly non  trivial. Indeed, in the case of four concentric cylinders we have many more states and the density of states is larger than that of a single cylinder having  the same length. In order
to explain better this point, let us  compute the density of states $\delta(k_B T )$ in a unit of thermal energy $k_B T$
for different numbers of dipoles $N$, see equation~(\ref{deltae-eq}) in the main text.
This is shown in Figure~\ref{pr4} 
  for both the concentric cylinders model and the MT, see Figure~\ref{pr4}(A).   As one can see the density of states is exactly the same   for the two models as a function of the number of dipoles $N$. Nevertheless  for the same fixed  length,
  the density of states for the four concentric cylinders is larger than the density of the single cylinder.  
Despite all that, a  large thermal delocalisation length for the concentric cylinder case can be explained by the fact 
that   the eigenstates  for the 4 concentric cylinders are delocalised  over a larger number of molecules as it is shown in Figure~\ref{pr4}(B).


\section*{References} 


\begin{thebibliography}{9}
\bibitem{dipsquare} Let us recall that, in Gaussian units, the unit dipole-dipole interaction energy is $[E] = [\mu]^2 [d]^{-3}$, where $[\mu]$ is the unit dipole and $[d]$ the unit distance. We express the dipoles in D (Debye), the distance in \AA \, and the energy in $\text{cm}^{-1}$ units (applying the standard conversion $[E]/(hc)$, with $h$ being the Planck constant and $c$ the speed of light), so that $[\mu]^2/(hc)=\text{cm}^{-1}\text{\AA}^3$. Now, from the definition $1 \mbox{ D}= 10^{-18} \mbox{ cm}^{5/2}\mbox{ g}^{1/2}\mbox{ s}^{-1}$ we have $1 \mbox{ D}^2= 10^{-12} \mbox{ cm}^2\mbox{ g}\mbox{ s}^{-2}\mbox{ \AA}^{3}$. Recalling the Planck constant $h = 6.626\cdot 10^{-27}\text{ cm}^2\text{ g s}^{-1}$ and the speed of light $c = 2.998\cdot 10^{10}\text{ cm s}^{-1}$, we have $1\text{ D}^2/(hc) = 5034\text{ cm}^{-1}\text{ \AA}^3$. So, a transition dipole $\mu = \sqrt{30}$~D results in $|\mu|^2 = 30 \times 5034\text{ cm}^{-1}\text{ \AA}^3 = 151020\text{ cm}^{-1}\text{ \AA}^3$. Note that in these calculations we write explicitly where the energy is divided by $hc$ for clarity, while in the main text we always assume implicitly that any energy is divided by $hc$.
\bibitem{Calculus} In terms of ESU-CGS and Gaussian units one may observe that $1 \mbox{ D}= 10^{-6} \mbox{ cm}\mbox{ g}^{1/2}\mbox{ s}^{-1}\mbox{ \AA}^{3/2}$ and the elementary charge is $|e^-|=4.80320425\cdot 10^{-10} \mbox{ cm}^{3/2} \mbox{ g}^{1/2} \mbox{ s}^{-1}$. From the relation $L_d=\mu/e^-$ we obtain $L_d=1.14 \mbox{ \AA}$.
\end{thebibliography}
\end{document}